\documentclass[12pt,a4p]{article}
\usepackage{epsfig}
\newlength{\picwi}
 \newcommand{\tcaption}[1]{
        \refstepcounter{table}
        \setbox\@tempboxa = \hbox{\footnotesize \bf Table~\thetable. #1}
        \ifdim \wd\@tempboxa > 6in
           {\begin{center}
        \parbox{6in}{\footnotesize\baselineskip=12pt \bf Table~\thetable. #1}
            \end{center}}
        \else
             {\begin{center}
             {\footnotesize \bf Table~\thetable. #1}
              \end{center}}
        \fi}
\begin{document}
%\setlength{\parskip}{\medskipamount}
%=======================================================================
\begin{titlepage}
\begin{center}{\large
    EUROPEAN ORGANIZATION FOR NUCLEAR RESEARCH
    }
\end{center}
 \bigskip
  \begin{flushright}
    CERN-EP-2000-070\\
    31 May 2000
  \end{flushright}
  \bigskip\bigskip\bigskip\bigskip\bigskip
\begin{center}
{\huge \bf  Multiplicities of $ \mathbf \pi^0,\ \eta,$
   $\rm \bf \ K^0\  $ and 
 of charged particles in quark and gluon jets \\ }
 \end{center}
 \bigskip\bigskip
 \begin{center}{\LARGE
      The OPAL Collaboration
      }
  \end{center}
  \bigskip\bigskip
\begin{abstract}
  We compared the multiplicities of $
  \pi^0, $ $ \eta, $  $\rm K^0 $ and of charged particles
  in quark and
  gluon jets in 3-jet events, as measured by the OPAL 
  experiment at LEP. The 
  comparisons were performed for distributions unfolded to $\rm 100\%$
  pure quark and gluon jets, at an 
  effective scale $\rm Q_{jet} $ which took into
  account topological dependences of the 3-jet environment.    
  The ratio of particle multiplicity in gluon jets to that in
  quark jets as a function of $\rm Q_{jet} $ for
  $ \pi^0,\ \eta$ or $\rm K^0 $
  was found to be independent of the particle species.
  This is consistent with the 
  QCD prediction that the observed enhancement in the mean particle rate
  in gluon jets with respect to quark jets should be independent of 
  particle species. 
  In contrast to some theoretical predictions and previous 
  observations, we observed no evidence for an
  enhancement of $ \eta $ meson production in gluon jets with respect
  to quark jets, beyond that observed for charged particles.
  We measured the ratio of the slope 
  of the average charged particle multiplicity 
  in gluon jets to that in quark jets, C, and we
  compared it to a next-to-next-to-next-to leading order 
  calculation. Our result, $\rm C=2.27\pm 0.20(stat+syst),$
  is about one standard deviation higher than the  
  perturbative prediction.        
\end{abstract}
 \bigskip
 \bigskip
 \begin{center}
%%% {\large \bf
%%%   This note describes preliminary OPAL results.
%%%   }   
 {\large \bf
   (Submitted to Eur. Phys. Jour. C)
   }  
 \end{center} 
\end{titlepage}

\begin{center}{\Large        The OPAL Collaboration
}\end{center}
\begin{center}
{\small
%begin authorlist PLEASE DO NOT DELETE THIS COMMENT
G.\thinspace Abbiendi$^{  2}$,
K.\thinspace Ackerstaff$^{  8}$,
C.\thinspace Ainsley$^{  5}$,
P.F.\thinspace Akesson$^{  3}$,
G.\thinspace Alexander$^{ 22}$,
J.\thinspace Allison$^{ 16}$,
K.J.\thinspace Anderson$^{  9}$,
S.\thinspace Arcelli$^{ 17}$,
S.\thinspace Asai$^{ 23}$,
S.F.\thinspace Ashby$^{  1}$
% ...
%end authorlist PLEASE DO NOT DELETE THIS COMMENT
} \end{center}\bigskip
\bigskip

{ \small
%begin institutes
$^{  1}$School of Physics and Astronomy, University of Birmingham,
Birmingham B15 2TT, UK
\newline
$^{  2}$Dipartimento di Fisica dell' Universit\`a di Bologna and INFN,
I-40126 Bologna, Italy
\newline
$^{  3}$Physikalisches Institut, Universit\"at Bonn,
D-53115 Bonn, Germany
\newline
$^{  4}$Department of Physics, University of California,
Riverside CA 92521, USA
\newline
$^{  5}$Cavendish Laboratory, Cambridge CB3 0HE, UK
\newline
$^{  6}$Ottawa-Carleton Institute for Physics,
Department of Physics, Carleton University,
Ottawa, Ontario K1S 5B6, Canada
\newline
$^{  7}$Centre for Research in Particle Physics,
Carleton University, Ottawa, Ontario K1S 5B6, Canada
\newline
$^{  8}$CERN, European Organisation for Nuclear Research,
CH-1211 Geneva 23, Switzerland
\newline
$^{  9}$Enrico Fermi Institute and Department of Physics,
University of Chicago, Chicago IL 60637, USA
\newline
$^{ 10}$Fakult\"at f\"ur Physik, Albert Ludwigs Universit\"at,
D-79104 Freiburg, Germany
\newline
$^{ 11}$Physikalisches Institut, Universit\"at
Heidelberg, D-69120 Heidelberg, Germany
\newline
$^{ 12}$Indiana University, Department of Physics,
Swain Hall West 117, Bloomington IN 47405, USA
\newline
$^{ 13}$Queen Mary and Westfield College, University of London,
London E1 4NS, UK
\newline
$^{ 14}$Technische Hochschule Aachen, III Physikalisches Institut,
Sommerfeldstrasse 26-28, D-52056 Aachen, Germany
\newline
$^{ 15}$University College London, London WC1E 6BT, UK
\newline
$^{ 16}$Department of Physics, Schuster Laboratory, The University,
Manchester M13 9PL, UK
\newline
$^{ 17}$Department of Physics, University of Maryland,
College Park, MD 20742, USA
\newline
$^{ 18}$Laboratoire de Physique Nucl\'eaire, Universit\'e de Montr\'eal,
Montr\'eal, Quebec H3C 3J7, Canada
\newline
$^{ 19}$University of Oregon, Department of Physics, Eugene
OR 97403, USA
\newline
$^{ 20}$CLRC Rutherford Appleton Laboratory, Chilton,
Didcot, Oxfordshire OX11 0QX, UK
\newline
$^{ 21}$Department of Physics, Technion-Israel Institute of
Technology, Haifa 32000, Israel
\newline
$^{ 22}$Department of Physics and Astronomy, Tel Aviv University,
Tel Aviv 69978, Israel
\newline
$^{ 23}$International Centre for Elementary Particle Physics and
Department of Physics, University of Tokyo, Tokyo 113-0033, and
Kobe University, Kobe 657-8501, Japan
\newline
$^{ 24}$Particle Physics Department, Weizmann Institute of Science,
Rehovot 76100, Israel
\newline
$^{ 25}$Universit\"at Hamburg/DESY, II Institut f\"ur Experimental
Physik, Notkestrasse 85, D-22607 Hamburg, Germany
\newline
$^{ 26}$University of Victoria, Department of Physics, P O Box 3055,
Victoria BC V8W 3P6, Canada
\newline
$^{ 27}$University of British Columbia, Department of Physics,
Vancouver BC V6T 1Z1, Canada
\newline
$^{ 28}$University of Alberta,  Department of Physics,
Edmonton AB T6G 2J1, Canada
\newline
$^{ 29}$Research Institute for Particle and Nuclear Physics,
H-1525 Budapest, P O  Box 49, Hungary
\newline
$^{ 30}$Institute of Nuclear Research,
H-4001 Debrecen, P O  Box 51, Hungary
\newline
$^{ 31}$Ludwigs-Maximilians-Universit\"at M\"unchen,
Sektion Physik, Am Coulombwall 1, D-85748 Garching, Germany
 }
\newline
%end institutes
\bigskip\newline
%begin notes
{ \small
$^{  a}$ and at TRIUMF, Vancouver, Canada V6T 2A3
\newline
$^{  b}$ and Royal Society University Research Fellow
\newline
$^{  c}$ and Institute of Nuclear Research, Debrecen, Hungary
\newline
$^{  d}$ and University of Mining and Metallurgy, Cracow
\newline
$^{  e}$ and Heisenberg Fellow
\newline
$^{  f}$ now at Yale University, Dept of Physics, New Haven, USA 
\newline
$^{  g}$ and Department of Experimental Physics, Lajos Kossuth University,
 Debrecen, Hungary
\newline
$^{  h}$ and MPI M\"unchen
\newline
$^{  i}$ now at MPI f\"ur Physik, 80805 M\"unchen
\newline
$^{  j}$ and Research Institute for Particle and Nuclear Physics,
Budapest, Hungary.
 }
%end notes

%======================================================================= 

\section{ Introduction}
  QCD predicts differences between quark and gluon jets. These are
  due to the different relative 
  probabilities for a gluon and a quark to radiate 
  an additional gluon, given by the SU(3) group constants
  $\rm C_A = 3 $ and $\rm C_F = 4/3.$ 
  The various measurements of these differences  
  from $\rm e^+e^- $ collider experiments 
  are found to agree with the theoretical predictions: gluon
  jets are observed to have higher mean particle multiplicity,
  a softer fragmentation function and a wider angular spread  than 
  light quark jets~\cite{qgresults}.\par 
  The QCD prediction of an enhancement of particle multiplicity in
  gluon jets with respect to quark jets is independent of
  the particle species, except for some small corrections.
%  due to non dynamical reasons. 
  This prediction can be tested by 
  measuring the rates of identified particles in gluon and quark jets. These 
  measurements  
  are also necessary for a better understanding of fragmentation 
  processes and hadronisation models. The models presently used are
  mainly the string~\cite{MC,jetset} and cluster 
  models~\cite{herwig} implemented in the Jetset~\cite{jetset}
  and Herwig~\cite{herwig} Monte Carlo 
  generators. Both models are based on the parton shower 
  approach, Jetset using a leading log perturbative QCD calculation, and
  Herwig    
  a next-to-leading log calculation.\par
  Some measurements of identified particle rates in quark and 
  gluon jets have already been made by the LEP experiments~\cite{LEPALL},
  and the ratios of rates in quark and gluon jets have been determined.
  The ratios of rates of $\rm K^+,\ K^0, $ and protons  
  were found to be consistent with the ratio of rates of the average 
  charged particle multiplicity 
  and also consistent with the ratio of rates
  determined from Monte Carlo simulations. The 
  ratio of rates for $\rm \Lambda $ 
  was measured by OPAL~\cite{LEPALL} and found to
  be larger than the Monte Carlo expectation. 
  The ratio of  the $ \eta $ meson production rate in 3-jet
  events to the $ \eta $ production rate in 2-jet events was 
  observed by the L3 experiment~\cite{L3ETA}  to be larger
  than the Monte Carlo expectation.
  It was suggested that this difference could be caused by an enhanced 
  $ \eta $ meson production in gluon jets.
  A confirmation of this result would
  suggest that, in addition to the QCD 
  predicted enhancement in gluon jets with respect to quark jets at
  equal jet energies, other $ \eta $ meson sources might exist in   
  gluon jets.  Such sources could be    
  production of glueballs and their decay to isoscalar mesons as has
  been foreseen in some theoretical models~\cite{Peterson}. However, 
  the ALEPH experiment~\cite{alepheta} observed no evidence for an 
  enhancement of $ \eta$ mesons 
  in gluon jets in excess of the Monte Carlo
  expectation.\par
  Experimentally, results from the comparison of charged particle or 
  identified particle rates in quark and gluon jets are highly 
  dependent on event topologies (i.e. the jet localisation with 
  respect to other jets in the event) and even more dependent on the 
  jet energies. Furthermore, comparison between the experimental 
  results and the QCD predictions is complicated by the use 
  of jet-finding algorithms. Analytic calculations do not employ    
  jet finders to assign particles to jets.
  To cope with these energy and topological dependences, a transverse 
  momentum-like scale $\rm Q_{jet} $ (see Equation 2)
  has been proposed~\cite{Qtheory} and was used 
  for a comparison of the mean charged
  particle multiplicity in quark and gluon jets~\cite{Delphi}.\par
  In this paper, the rates of  $ \pi^0,\ \eta $ and   
  $\rm  K^0 $ in quark and gluon  
  jets are measured for the first time
  as a function of the scale $\rm Q_{jet} $.
  A phenomenological formula for the charged particle rate
  $\rm N_q(Q_{jet}) $ in quark jets,
  written as a second order polynomial in $\rm \ln(Q_{jet}), $ 
  and a phenomenological formula for the charged particle rate
  $\rm N_g(Q_{jet}) $ in gluon jets, written as a
  linear function of $\rm N_q(Q_{jet}), $ 
  were  used to describe simultaneously the observed charged
  particle multiplicities in quark and in gluon jets. 
  The functions $\rm N_q(Q_{jet}) $ and $\rm N_g(Q_{jet}) $, with 
  all parameters
  set to the values obtained from a fit to the observed charged particle
  distributions, 
  were used as a model to compare to the measured rates of identified
  particles in quark and gluon jets. For each comparison, only one
  overall normalisation factor, common to $\rm N_q(Q_{jet}) $ and 
  $\rm N_g(Q_{jet}), $ is allowed to vary.
  This method provides a model independent way of testing that the 
  enhancement of particle production in gluon jets with respect 
  to quark jets is independent of the particle species.\par
\section{Data selection }
\subsection{Selection of hadronic $\rm \mathbf{Z^0} $ decays}
 The present analysis was based on the full hadronic event sample 
 collected at centre-of-mass energies at and near the $\rm Z^0 $ peak by the 
 OPAL detector from 1991 to 1995. This corresponded to about 4 million 
 hadronic $\rm Z^0 $ decays.
 A full description of the OPAL detector can be found in~\cite{opald}. 
 Standard OPAL selection criteria were  applied for track and
 electromagnetic cluster selection~\cite{billg}. 
 Tracks were required to have:  at least
 20 measured points in the jet chamber, 
 a measured
 momentum greater than $\rm 0.10\ GeV $, an impact 
 parameter $\rm \mid d_0\mid $
 in the $\rm r-\phi $ plane smaller than 2  cm, a z position within 
 25 cm 
 of the interaction point and a measured angle with respect to the beam axis
 of greater than $ \rm 20^\circ .$
 Electromagnetic
 clusters were required to have an energy greater 
 than $\rm  0.1\ GeV $ if they 
 were in the barrel part of the detector (i.e.  
 $\rm \mid \cos{\theta}\mid\ \le 0.82 $) 
 or greater than $\rm 0.3\ GeV $
 if they were in the endcap part. The selected tracks and
 clusters not associated with tracks 
 were fed, as four-vectors, to the jet-finding algorithms.
 Background from all sources
 was reduced to less than $\rm 0.8 \% $ and was neglected
 throughout the analysis. It was reduced 
 by requiring for each event more than 7 measured tracks, a visible 
 energy (i.e. the sum of detected particle
 energies after correcting for 
 double counting) larger than 60 $\rm GeV $ and an angle
larger than 
 $\rm 25^\circ$ between the calculated thrust axis and the beam axis .\par
\subsection{ Event simulation }
 Detector effects and detection efficiencies for the studied
 particles were evaluated using 8 million Monte Carlo hadronic
 $\rm Z^0 $ decays. Events were generated using the Jetset program tuned to
 reproduce the global features of hadronic events as measured 
 with the OPAL detector. About 4 million events generated by
 the Herwig program were also used for comparison.
 The generated events were processed through a full simulation of the 
 OPAL detector~\cite{opalsim} and were processed using the same 
 reconstruction and selection
 programs as were applied to the data.\par
\subsection{Selection of 3-jet events}
   Three jet-finding algorithms were used: Luclus~\cite{jetset}, 
   Durham~\cite{durham} and the cone~\cite{cone} jet finder.
   The Luclus jet finder was found to provide the best jet angular 
   resolution, which was relevant for the present analysis. Luclus was  
   therefore used as the reference algorithm
   while the two others were used for systematic comparisons.
   The jet algorithm was forced to resolve three jets in each hadronic
   event. The jet energies and momenta were then calculated by 
   imposing energy and momentum conservation with planar 
   massless kinematics~\cite{massless},
   using the jet directions found by the 
   jet algorithm. They are given by the cyclic relation:
  \begin{equation}
    \rm E_i = \frac{\sqrt{s}\cdot\sin{\theta_{j,k}}}
    {\sin{\theta_{i,j}}+\sin{\theta_{j,k}}+\sin{\theta_{k,i}}},
  \end{equation}
  where $\rm \theta_{i,j} $ is the angle between jet i and jet j.
   The event was accepted as a 3-jet event if each jet contained 
   at least 3 charged particles, had a corrected energy exceeding
   $\rm 5\ GeV$, and pointed more than $\rm 20^\circ $ away 
   from the beam axis and more 
   than $\rm 30^\circ $ away from the direction of the other two jets.\par
   The variable 
   $\rm Y=(D_{2\rightarrow 3}- D_{3\rightarrow 4})/E_{visible} $
   was used, where
   $\rm D_{2\rightarrow 3}\ and \ D_{3\rightarrow 4} $ were 
   the Luclus jet algorithm 
   resolution parameters ($\rm D_{join}$)~\cite{jetset}
   for the transition from 2 to 3 and   
   3 to 4 jets, respectively, and  $\rm E_{visible} $ the total visible 
   energy.  This variable  
   measured the topological stability of the event 
   as a 3-jet event. For larger values of Y, the events 
   tended to be
   three-fold symmetric, meaning that all inter-jet 
   angles were close to $\rm 120^\circ.$ 
   To select stable 3-jet events (i.e. events not close to the
   transition from three to four jets), only 
   events with $\rm Y>0.2 $ \footnote{The corresponding Y variable for Durham 
   was defined as $\rm  Y=y_{2\rightarrow 3}- y_{3\rightarrow 4}$, where 
   y is the usual $\rm y_{cut}$~\cite{durham}.  
   For the cone jet finder the inter-jet
   angles were considered instead.} 
   were kept for further processing.  \par
   The total selected data sample contained 
   approximately 493 000 3-jet events,
   which was  $\rm 12.32\% $ of the total initial event sample. The 
   corresponding fraction for Monte Carlo events was 
   $\rm 12.30\% $ for Jetset and  $\rm 12.17\% $ for Herwig.\par
   For the Monte Carlo events, the jet-finding algorithm was applied at
   the parton, hadron and detector levels (defined in Section 3.2).
   At each level the jet energies 
   were corrected to satisfy the constraints of energy 
   and momentum conservation
   with planar massless kinematics, after
   which the jets were energy ordered, the first jet being the jet with 
   the highest energy. 
   The matching from parton to hadron level and
   from hadron to detector level was done using a simple 
   angular correspondence: a jet at hadron level
   was matched with only one detector jet and one parton jet, those
   having the minimal 
   angular deviation with respect to the hadron jet direction.    
   The jet energy resolution, defined as 
   $\rm (E_{jet}^{parton} - E_{jet}^{detector})/E_{jet}^{parton}, $ was found 
   to range from $\rm 5\% $ for the first jet  to $\rm 13\% $ for the third 
   jet.
   The angle between 
   the parton jet direction and the detector jet direction was found 
   to have an r.m.s. of $\rm 0.07\  radians $ for the highest 
   energy jet and 
   $\rm 0.16\ radians $ for the lowest energy jet.
\section{ Analysis method }
   In this section, the jet scale $\rm Q_{jet} $ and
   the jet purities are defined and the method of unfolding the 
   average charged particle multiplicity to
   $\rm 100\% $ pure quark and gluon jets is explained. In
   Section 4 the measured
   average charged particle multiplicity in pure quark and gluon jets, as
   a function of the scale $\rm Q_{jet} $, was fitted to   
   phenomenological formulae.
   The purpose of the fit was to obtain an analytical 
   shape that could be used 
   as a reference or a model to compare in Section 6
   to the corresponding shapes 
   obtained in Section 5 for $ \pi^0,\  \eta$ and
   $\rm K^0 $.
   Systematic effects that could have distorted the measured shape are
   discussed in Section 4.3.\par         
\subsection{ Jet scale $\rm Q_{jet} $} 
   It has already been shown~\cite{Qtheory}
   that the jet energy alone is not an adequate scale to describe the average
   particle multiplicity in quark and gluon jets.   
   Coherence in QCD radiation
   suggests~\cite{Qtheory2} that the  position of a  
   parton with respect to other partons in the event (i.e.
   the event topology) should also be considered in studies 
   of jet properties. Inter-jet coherence effects that can lead to
   destructive or constructive interference effects on the 
   particle flux in the inter-jet angular region, have been predicted
   and observed experimentally. An example is the 
   string effect~\cite{string}. 
   A transverse-momentum-like scale combining the jet energy
   and its angular position with respect to the other jets
   has been used~\cite{Qtheory2} in a phenomenological 
   study of   
   parton shower characteristics. This scale $\rm Q_{jet} $ was defined as:
\begin{equation}
   \rm Q_{jet} = E_{jet} \sin({\theta/2)} 
\end{equation}
   where $\rm \theta $ was the jet angle with respect to the closest jet.   
   $\rm Q_{jet} $ has already been used
   in an experimental study of multiplicity 
   in $\rm e^+e^- $ 3-jet events~\cite{Delphi}.
% for a determination of 
% $\rm C_A/C_F $ for events with a particular topology.
   The way the scale $\rm Q_{jet} $ incorporates 
   the topological dependence can 
   be seen, for example, at leading order in $\rm \alpha_s $ 
   for the three parton $\rm q\bar{q}g $ vertex: colour being conserved
   in QCD, the gluon can be represented by a $\rm q\bar{q} $ pair 
   that compensates exactly the total colour charge of the initial 
   $\rm q\bar{q} $ pair. A mutual colour shielding occurs when the initial 
   quark or anti-quark is close in angle to the gluon 
   and reduces subsequent gluon radiation.
   This colour shielding is minimal for back-to-back partons, and the
   scale $\rm Q_{jet} $ becomes equal 
   to the parton energy for that case.\par
\begin{figure}
        \centerline{\includegraphics[scale=0.8]{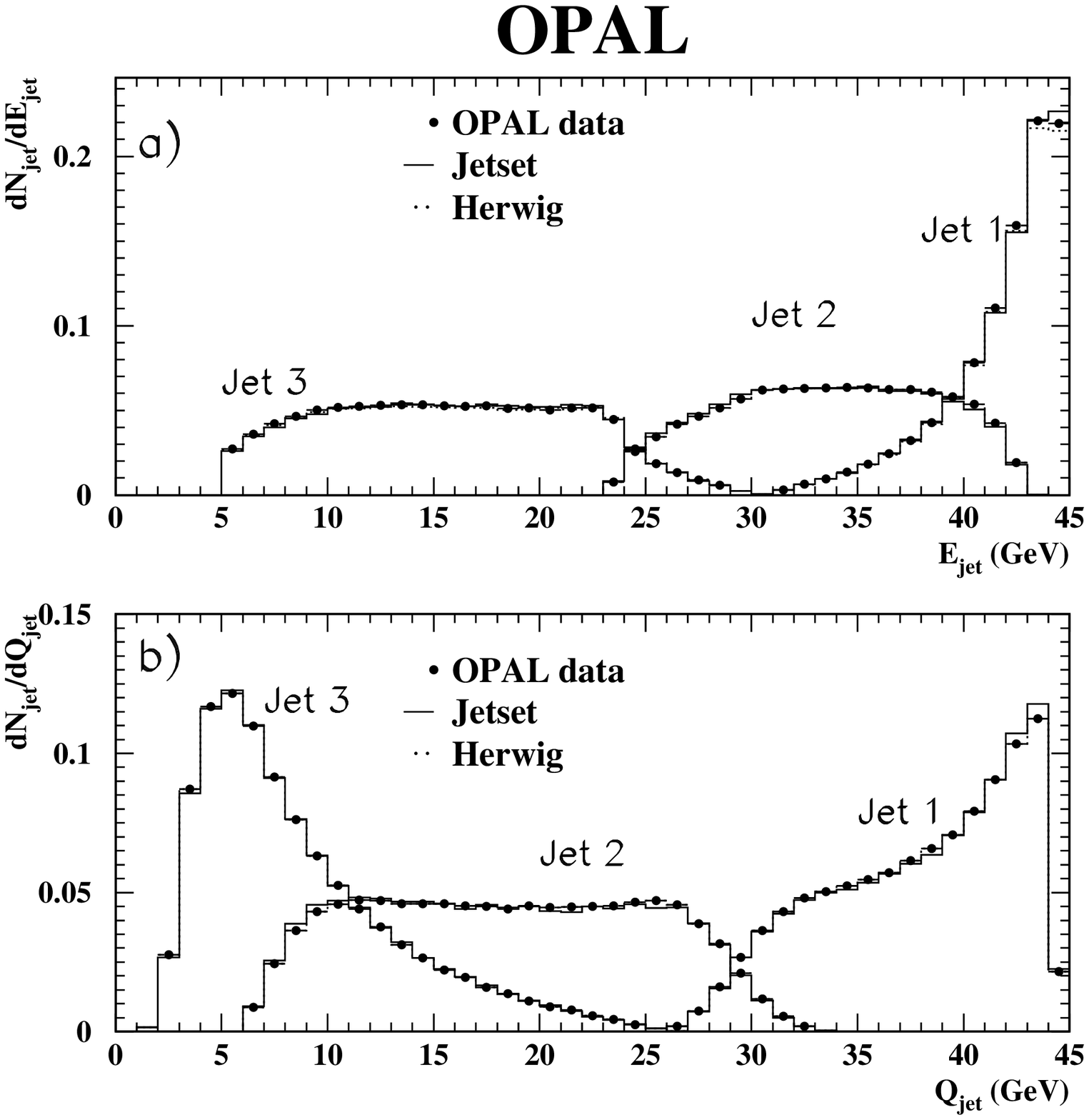}}
        \caption{ a) 
         Jet energy ($\rm E_{jet1}>E_{jet2}>E_{jet3} $)  and b) jet scale 
          $\rm Q_{jet} = E_{jet} \sin({\theta/2}) $ distributions for 
          the selected events.}           
        \label{fig:qscale}
\end{figure}

   The scale $\rm Q_{jet} $ was used for the present 
   analysis with no restriction on event topology except for 
   the minimal inter-jet angle. 
   The distributions of energy, $\rm E_{jet},$ and scale, $\rm Q_{jet} $ 
  (normalised to the total number of analysed hadronic events), for 
   the three energy ordered jets, 
   are shown in Figures \ref{fig:qscale}a and \ref{fig:qscale}b.
   The Monte Carlo distributions were found to reproduce the data
   very well.\par
\subsection{ Gluon jet definition and purity estimation}
   The jet having the 
   smallest energy, $\rm E_{jet},$ was found to have the 
   highest gluon purity.  
   The jet purities were estimated using the Monte
   Carlo information: the initial parton shower (parton level),
   the generated hadrons after the fragmentation processes (hadron level)
   and the reconstructed particles after the simulation
   of the full OPAL detector response (detector level). At the hadron level,
   all charged and neutral particles with lifetimes greater than 
   $\rm 3\times 10^{-10}s $ were considered as stable particles.     
   A jet at the detector level was considered to be a gluon jet if it 
   matched
   a parton jet that did not contain either the initial quark
   or anti-quark from the Z decay.
   The jet purities were determined directly from Monte Carlo as
   the fraction of quark or gluon jets  
   present in the jet sample at a fixed scale $\rm Q_{jet} $.\par
\begin{figure}
        \centerline{\includegraphics[scale=0.8]{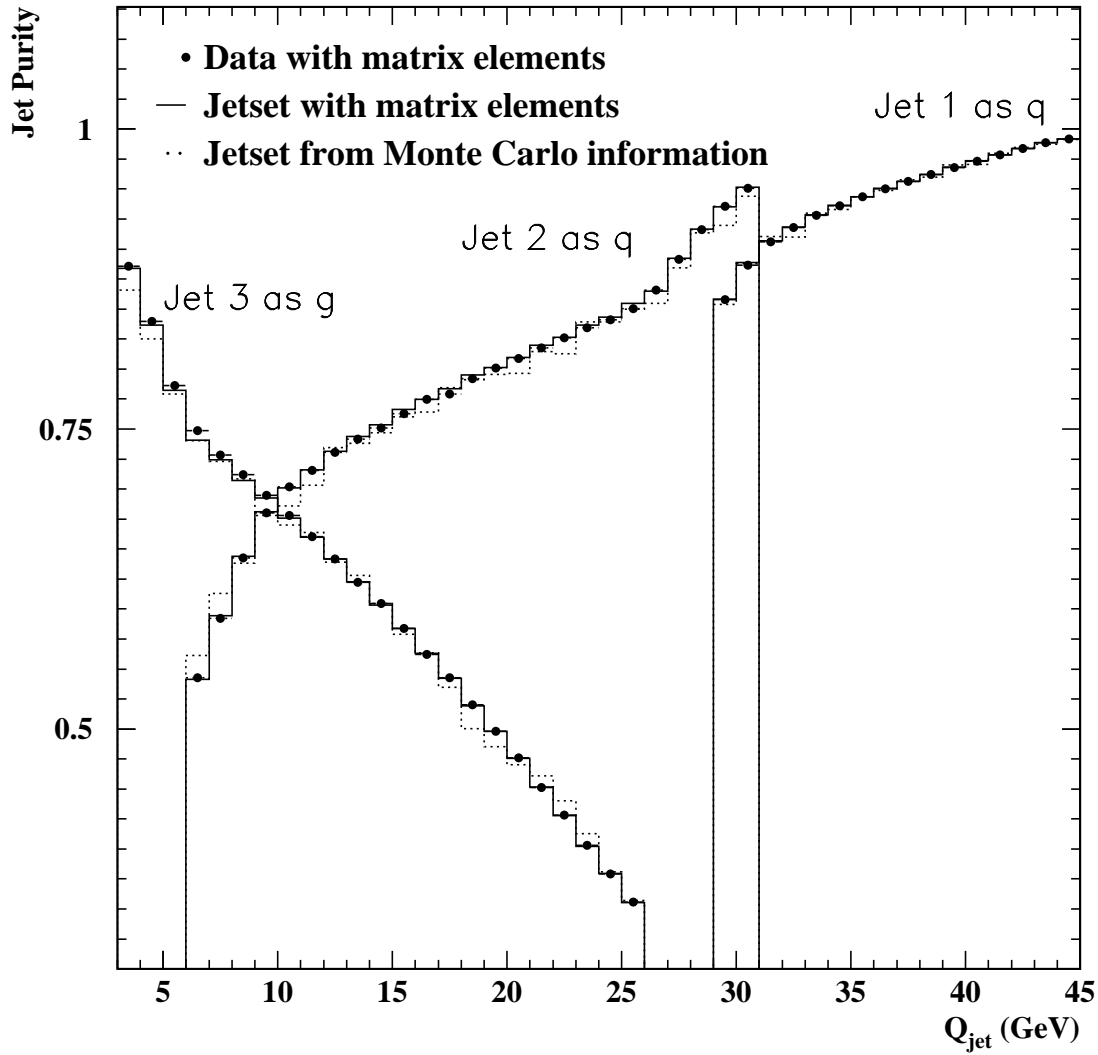}}
        \caption{ Quark jet purities of the first    
        and second jet, and gluon jet purity of the third  
        jet as a function of the scale $\rm Q_{jet} $. Jets are
        energy ordered. 
        The purities of OPAL data
        obtained from the matrix element formula are shown together with 
        the purities for Jetset Monte Carlo events estimated from 
        Monte Carlo information and from the matrix element
        formula.}
        \label{fig:purities}
\end{figure}

   The purities could also be estimated from matrix element calculations. It 
   has been shown~\cite{MX} that, for leading order QCD matrix 
   elements, 
   the probability for a given jet 
    $\rm \{i\} $ among the jets $\rm \{i,j,k\} $
   to be a gluon jet can be expressed as
   a function of the jet energies:
\begin{equation}
   \rm  P_{i=g} \propto \frac{x_j^2+x_k^2}{(1-x_j)(1-x_k)}, 
\end{equation}
   where $\rm x_i= 2E_i/\sqrt{s} $ and the corresponding probability
   for being a quark jet is
\begin{equation} 
    \rm P_{i=q} = 1-P_{i=g},
\end{equation} 
    normalised to have:
\begin{equation}
    \rm P_{1=q}+P_{2=q}+P_{3=q} = 2.
\end{equation}
%   The probabilities were normalised 
%   by assuming that there was one gluon jet per selected 3-jet event.
   The purities obtained using Monte Carlo information are shown
   in Figure \ref{fig:purities} together with the purities obtained 
   from the matrix elements.  In the same figure are also shown 
   the purities for OPAL data obtained from the matrix element formula. 
   Very good agreement
   was obtained between the two methods, with less than $\rm 2\% $ 
   deviation in most of the range of $\rm Q_{jet} $ common to 
   the second and third jet (see Figure 1 and Figure 2).
   Therefore, the matrix element formula was used to estimate 
   the jet purities directly from the data. 
   The same agreement was also observed 
   for Herwig Monte Carlo events.\par
\subsection{Unfolding to pure quark and gluon jets}      
    The measured jet samples were mixtures of quark and gluon jets, while
   meaningful comparison of gluon and quark jets should 
   be performed on quantities evaluated for pure samples of gluon jets and pure
   samples of quark jets. This analysis   
   used the average charged particle 
   multiplicity for samples of jets at the same scale $\rm Q_{jet} $
   but having  
   different gluon/quark jet purities.
   If $\rm \langle N_1(Q_{jet})\rangle $ and   $\rm \langle N_2(Q_{jet})
    \rangle $ 
   are the average 
   measured charged particle multiplicities for two jet samples with different 
   quark jet purities $\rm P_1(Q_{jet}) $ and $\rm P_2(Q_{jet}) $ 
   at the same $\rm Q_{jet} $ value, then:
\begin{equation}
    \rm  \langle N_1(Q_{jet})\rangle  = P_1(Q_{jet})\langle 
      N_q(Q_{jet})\rangle + 
       (1-P_1(Q_{jet}))\langle N_g(Q_{jet})\rangle
\end{equation}
%      The total rates were corrected for the inaccessible $\rm x_E $
%      regions. The corrections were performed using extrapolation with  
%      the spectral shapes predicted by Jetset. 
%      They amounted to 
%     $\rm 11\%,\ 23\% $  and $\rm 0.6 \% $ of the total rates for 
%     $ \pi^0,\ \eta $  and $\rm K^0 $ respectively.
\begin{equation}
    \rm  \langle N_2(Q_{jet})\rangle = P_2(Q_{jet})\langle N_q(Q_{jet})
        \rangle + 
       (1-P_2(Q_{jet}))\langle N_g(Q_{jet})\rangle
\end{equation}
   where $\rm \langle N_q(Q_{jet})\rangle\ and \ 
        \langle N_g(Q_{jet})\rangle  $ are the average particle 
   multiplicities for $\rm 100 \% $ pure quark and
   gluon jet sample. This gives:
\begin{equation}
   \rm N_q(Q_{jet}) = \frac{(1-P_2(Q_{jet}))\langle N_1(Q_{jet})\rangle+
      (1-P_1(Q_{jet}))\langle N_2(Q_{jet})\rangle }
               {P_1(Q_{jet})-P_2(Q_{jet})} 
\end{equation}
\begin{equation}
   \rm N_g(Q_{jet}) = \frac{P_1(Q_{jet})\langle N_2(Q_{jet})\rangle-
       P_2(Q_{jet})\langle N_1(Q_{jet})\rangle }
               {P_1(Q_{jet})-P_2(Q_{jet})}.\quad 
\end{equation} 
    This unfolding was only possible in the region where
    the $\rm Q_{jet} $ scales of the jet samples 
    overlapped.
    The $\rm Q_{jet} $  distributions of the 
    energy ordered jets (see Figure \ref{fig:qscale}) showed that the 
    second and the third jets could be considered  
    to be different samples over a common range of $\rm Q_{jet} $  from 
    $\rm 6\  to\  26\ GeV$.\par
\section{Average charged particle multiplicities }
\subsection{Measurements and parametrisation}  
\begin{figure}
       \centerline{\includegraphics[scale=0.8]{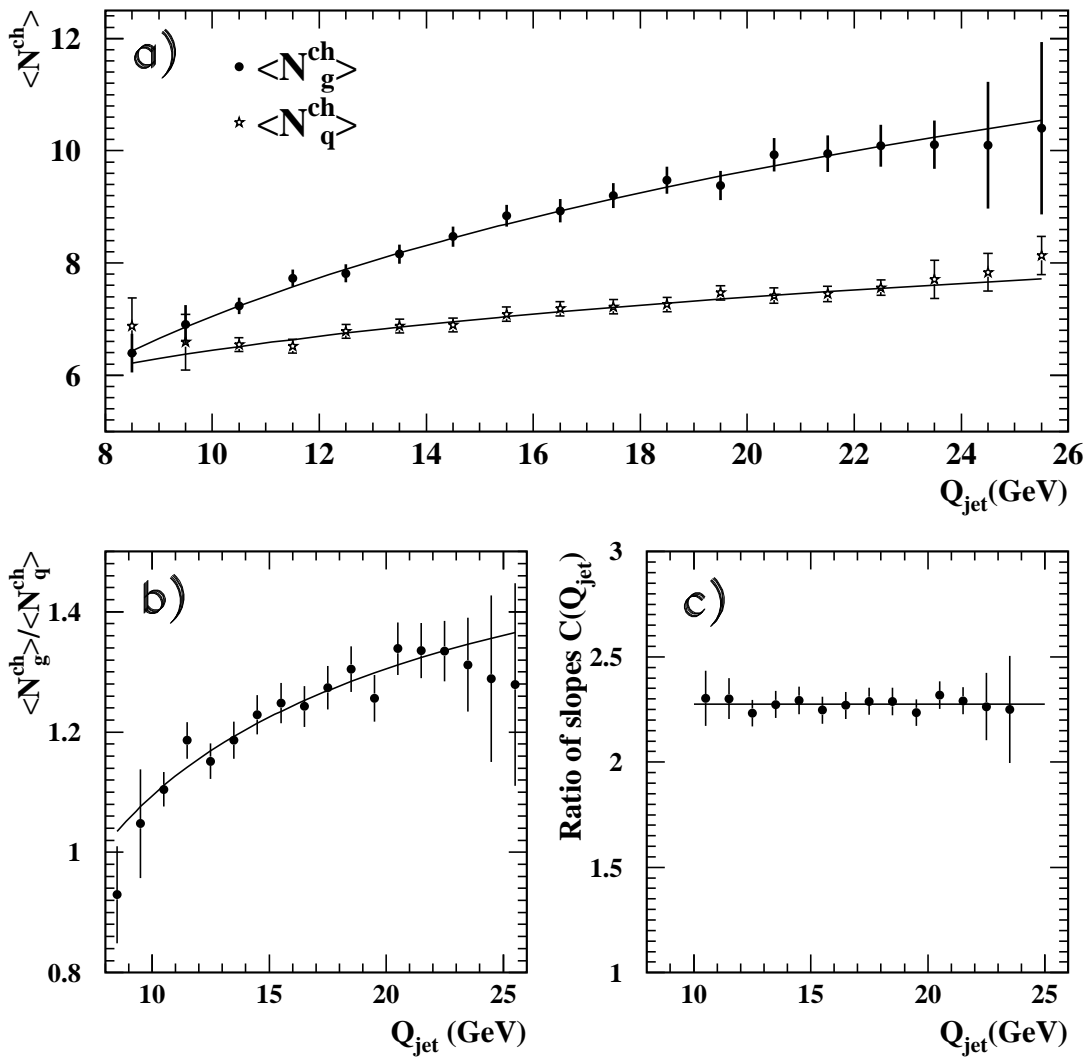}}
       \caption{ a) Average number of charged particles for  
         pure gluon and quark jets as a function of the scale $\rm Q_{jet} $.
         Systematic errors are included.
         The continuous curves are from a fit to a  
         phenomenological formula (see text).
         b) Ratio of the average charged 
            particle multiplicities in gluon and
            quark jets as a function of the scale $\rm Q_{jet} $.
         c) Ratio of the average charged
            particle multiplicity slopes in gluon and quark jets;
            the line is a fit to a constant.} 
       \label{fig:charge}
\end{figure}

  The average number of charged particles, 
  $\rm \langle N^{ch} \rangle, $ per jet in bins of 1 $\rm GeV $
  of the scale $\rm Q_{jet} $  was measured for samples
  of the second and third jets, being respectively 
  quark enriched and gluon enriched through
  jet energy ordering and having a wide common range of 
  $\rm Q_{jet}$.
  In each bin of $\rm Q_{jet} $, the average purity was obtained from 
  the matrix element formula. 
   The efficiency corrections from detector to
   hadron level were calculated from Monte
   Carlo information as a function of $\rm Q_{jet} $ for each
   jet sample separately. The efficiency was defined as the ratio 
   of the average
   number of charged particles at the detector level
   for a given jet sample at a given
   scale $\rm Q_{jet} $, divided by 
   the equivalent quantity for the corresponding jet sample  
   at the hadron level. The efficiencies were found 
   to be approximately independent of $\rm Q_{jet} $ for all jet samples, 
   and the corresponding 
   corrections to $\rm \langle N^{ch}(Q_{jet}) \rangle $ 
    were at the level of $\rm 10\%$.
   The results obtained after unfolding to $\rm 100\% $ pure
   quark and gluon jets are shown in Figure \ref{fig:charge}. The 
   unfolding was performed for $\rm Q_{jet} > 7\ GeV, $
   since the method worked for all three   
   jet finders in this region.\par 
   The average charged particle multiplicities in quark jets, $\rm 
   \langle N_q \rangle,  $ 
   and in gluon jets,  $\rm \langle N_g \rangle,  $
   were simultaneously described with phenomenological formulae. 
   The  parametrisation was given by:
\begin{equation}
   \rm \langle N_q(Q_{jet})\rangle 
   = a_0 + a_1 \ln{Q_{jet}} + a_2  (\ln{Q_{jet}})^2 
\end{equation}
\begin{equation}
   \rm \langle N_g(Q_{jet})\rangle =  R_0 + R_1 \langle N_q(Q_{jet})\rangle 
\end{equation}
   where $\rm a_0,\ a_1,\ a_2,\ R_0\ and \ R_1 $ are constants.
   This parametrisation resembles the next-to-leading order
   expressions given in~\cite{gaffney,Webber}, but with an
   extra offset parameter $\rm R_0. $  The leading order 
   expressions~\cite{gaffney,Webber} alone were found
   to be unable to fit the data, as was also found in~\cite{Delphi} and
   discussed in~\cite{JWG}.
   The $\rm Q_{jet} $ dependent 
   average charged particle multiplicities for quark and gluon jets
   were fitted simultaneously to the
   expressions for $\rm  \langle N_1(Q_{jet}) \rangle \ and \
   \langle N_2(Q_{jet})\rangle $ (Equations 6 and 7) where
   $\rm  \langle N_q(Q_{jet}) \rangle \ and \langle N_q(Q_{jet})\rangle $
   were replaced by the expressions given in equations 10 and 11.
   The result of the fit yielded:
    $\rm a_0=2.74\pm 0.07,\ a_1=1.71\pm 0.05,\ a_2=-0.05
     \pm 0.07 ,\ R_0=-7.27\pm 0.52\ and\ R_1=2.27\pm 0.07. $
   The fitted functions are shown with the 
   data points unfolded using equations 8 and 9
   in Figure \ref{fig:charge}a, where a good description 
   of the average charged particle multiplicities in
   both the gluon and the quark jets can be seen.\par 
   The ratio of the average 
   charged particle multiplicities in gluon to quark jets is shown
   in Figure \ref{fig:charge}b. This ratio is predicted by QCD to be
   the same for all particle species. 
   A test of this prediction is        
   the quality of the fits of the above analytical function for charged
   particles to the  measured multiplicities of
   $ \pi^0,\ \eta$ and $\rm K^0 $ in 
   quark and gluon jets with all parameters fixed except for an overall
   normalisation factor.\par
\subsection{Ratio of the slopes of the multiplicities} 
   From QCD perturbative calculations, the ratio of particle multiplicities in
   gluon and quark jets is given by:
\begin{equation}
   \rm N_{g}(Q_{jet}) = R(Q_{jet})\times N_{q}(Q_{jet})
\end{equation} 
   where the asymptotic limit of $\rm R(Q_{jet}) $
   at large $\rm Q_{jet} $
   is an approximation to the QCD colour factor ratio
   $\rm R=C_A/C_F=2.25. $ At the $\rm Z^0 $ scale, this value is 
   lower due to sizeable higher order corrections. However,
   the ratio of the slopes
\begin{equation} 
   \rm C(Q_{jet})={d\langle N_{g}\rangle /dQ_{jet}
     \over{d\langle N_{q} \rangle /dQ_{jet}}}(Q_{jet}) 
\end{equation}
   of multiplicities in quark and gluon jets is expected to be 
   less affected by these higher order corrections~\cite{Capella}.
   The slope ratio  C has been recently calculated~\cite{Capella}
   using a 
   next-to-next-to-next-to leading order (3NLO) perturbative approximation.
   The predicted value of $\rm C(Q_{jet}) $ in
   $\rm  Z^0 $  decays
   is $\rm C\simeq 1.92. $\par
   From the parametrisation given in equations 10 and 11 
   that imposes a constant value for $\rm C(Q_{jet})$ (Equation 13),   
   we obtained:
  \begin{equation}
    \rm C(Q_{jet})= R_1 = 2.27\pm 0.07(stat.)\pm 0.19(syst.). 
  \end{equation} 
   The constraint $\rm C(Q_{jet})= constant $ can be released
   by extracting the spectrum $\rm C(Q_{jet})$  
   from the measured distributions  $ \rm \langle N_q(Q_{jet})\rangle $
   and $\rm  \langle N_q(Q_{jet})\rangle.$   
   This was done by using as an estimate of the derivative at 
   each bin $\rm Q_{jet} $ the slope of 
   a fitted line to three adjacent bins centred at $\rm Q_{jet}. $ 
   The obtained $\rm C(Q_{jet})$ spectrum
   is shown in Figure \ref{fig:charge}c, and
   a fit to a constant yielded:
   $$\rm C = 2.27 \pm 0.09(stat) \pm 0.27(syst). $$ 
   The systematic uncertainty, for both methods, was 
   mainly due to differences between the jet finders and the correlations
   between the unfolded values of $\rm N_{g}(Q_{jet}) $ and 
    $\rm N_{q}(Q_{jet}) $
   at each bin of $\rm Q_{jet} $. 
   The two values are about one standard deviation higher than
   the prediction~\cite{Capella} $\rm C\simeq 1.92.$\par  
   The Delphi collaboration recently presented a 
   measurement of the ratio of slopes\cite{Delphi}.
   Their result,   
   $\rm C=1.97\pm 0.10(stat), $ is about one standard deviation of the
   total uncertainty bellow our measurement.\par
%   which is consistent with our result.\par  
\subsection{Stability of the parametrisation of multiplicities} 
 To study systematic effects on the parametrisation 
 of the charged particle multiplicity obtained in the previous
 section, several variations to the analysis procedure were
 applied. For each variation, new  $ \rm \langle N_q(Q_{jet})\rangle $
 and $\rm \langle N_q(Q_{jet})\rangle$ were obtained and compared
 to the original distributions. At each data point, 
 the deviation with respect to the original value
 is  considered as a systematic
 error. The errors were added quadratically and were included
 in the error bars of the original 
 data points shown in Figure \ref{fig:charge}.
 For each variation, the distributions $ \rm \langle N_q(Q_{jet})\rangle $
 and $\rm \langle N_q(Q_{jet})\rangle$ were also 
 fitted to the functions of equations 10 and 11. The
 resulting parametrisation was compared to 
 the original one by calculating the largest relative difference
 between the new and the original function values.
 The following  systematic variations were considered:
\begin{enumerate}  
 \item
  The analysis was completely repeated  with the Durham and cone jet finders. 
  The parametrisation was found to agree well 
  between the Luclus (original) and the Durham jet finders.
  The  parametrisation obtained using the cone jet finder agreed within
  $\rm 10\% $ with the original one.
%%% in that
%  $\rm 10\% $ was the largest relative difference 
%  between the new and the original parametrisation                   
%
%  The analytical shape was also checked for
%  stability against the considerations listed below, 
%
 \item The analysis was repeated using a different method for estimating 
       the efficiencies. The fully unfolded 
       average charged particle multiplicities were obtained, 
       as a function of $\rm Q_{jet} $
        and for each jet sample,  at the hadron level
       and then at the detector level. The ratio of detector to hadron level
       was then used to correct the average multiplicities of the data 
       after having performed the unfolding. This procedure was done 
       with the Jetset and Herwig Monte Carlos and the  
       corrections were found to differ by at most 
       $\rm 3\%. $  The analytic formula (Equations 10 and 11) 
       was then fitted to 
       the average charged particle multiplicity spectra corrected separately 
       with Jetset and Herwig. The 
       shape was found to agree with the original one within 
       $\rm 2\% $ for Jetset and $\rm 3\% $ for Herwig. 
 \item The influence of the jet purities was studied 
       by using the purities from the Monte Carlo matching,
       and also by increasing the 
       cut on D from 0.20 to 0.25. The effect on the
       fitted parametrisation was negligible.
 \item Effects of soft particles on the measured multiplicities
       were studied by  changing 
       the minimum momentum required 
       per track from 0.10 GeV to $\rm 0.15\ GeV,$ 
       and the  minimum number of tracks per jet was also changed from 
       3 to 5. The analysis was repeated and the fitted parametrisation
       agreed with the original one within $\rm 3\%.$
 \item
     The analysis was repeated with            
     two different jet samples. 
     The first jet sample was gluon enriched by requiring that
     two jets were tagged as b-quark
     jets with a neural network b-tagging method and 
     the remaining jet was considered as a gluon jet.    
     The second jet sample, which was quark enriched,
     was  obtained by selecting 
     all second jets (energy ordered) 
     in the events having no b-tagged quark jets.
     The average
     gluon purity was $\rm 80\% $ for the first jet sample,
     and the average quark purity was $\rm  65\% $ 
     for the second jet sample.
     The fitted parametrisation agreed with the original one within 
     $\rm 8\%.$
     The new average charged particle multiplicity spectra,
     unfolded to  $\rm 100\% $ quark and gluon purities,
     agreed very well with the original spectra. Differences
     at each data point were considered as systematic errors; this
     should, in principle, also account for possible correlations between
     the measured 
     $\rm N_g(Q_{jet})\ and \ N_q(Q_{jet}).$ 

\end{enumerate} 
\section{$ \pi^0,\ \eta $ and $\rm K^0 $ meson production }
This section describes the reconstruction of the decay channels 
$ \eta\rightarrow 2\gamma,\ \pi^0\rightarrow 2\gamma $ and
$\rm K^0_S $ $ \rightarrow \pi^+\pi^- $
for the full $\rm Z^0 $ hadronic decay sample without 
any 3-jet requirement.
In order to gain more confidence for the rate measurements in 
quark and gluon jets (see Section 6), 
the total inclusive rates as well as the differential rates in 
hadronic $\rm Z^0 $ decays
of $ \pi^0,\ \eta $ and $\rm K^0 $ were measured 
and compared to previous such measurements at LEP. 
\subsection{$ \pi^0 $ and  $ \eta $ reconstruction} 
\begin{figure}
        \centerline{\includegraphics[scale=0.8]{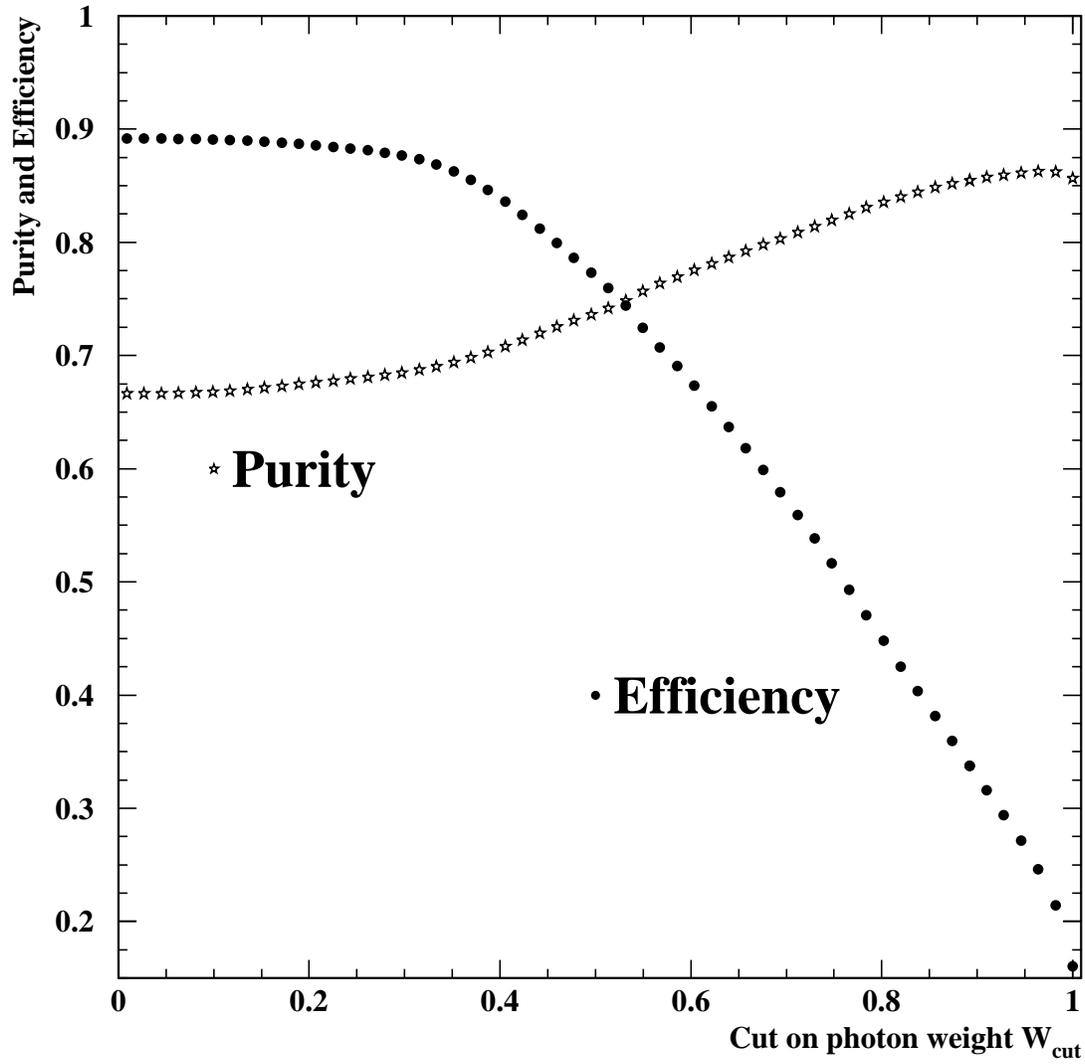}}
        \caption{Purity and efficiency of the photon reconstruction versus  
         the cut value on the associated weight W. } 
        \label{fig:Purity}
\end{figure}

The reconstruction was restricted to the barrel region of the detector.
A procedure~\cite{TN527} using a parametrisation of the
expected lateral energy distribution of electromagnetic showers was
optimised to resolve as many photon candidates as possible from the
overlapping energy deposits in the electromagnetic calorimeter, in 
the dense environment of hadronic jets.
The procedure was efficient but led to a rather low purity due 
to reconstructing spurious photons (``fakes'').
Based on Monte Carlo simulated events, the rejection of fake photons
was studied using a set of five measurable variables, namely:
$\rm E_{\gamma}:$ the energy of the photon candidate,
$\rm E_{clust}: $  the energy
of the nearest cluster to the considered photon candidate,
$\rm \theta_{clust}: $ the opening angle between the 
photon candidate and the nearest cluster,
$\rm \theta_{trk}: $  the opening angle between the 
photon candidate and the closest reconstructed track,
and $\rm E_{trk}: $  the amount of energy that could be
attributed to tracks in an array of 3x3 lead glass blocks
around the position of the photon candidate. 
A large number of obvious fake photons 
(mostly with $\rm E_{\gamma} \leq 300$ $ \rm MeV $)
were rejected with the following two cuts:  
 \begin{enumerate}
  \item  A candidate was rejected if it was found
          to satisfy:
\begin{equation}
          \rm E_{\gamma} < A\cdot E_{clust}\cdot 
             exp\biggl(- \left(\frac{\theta_{clust}}{40\ mrad}\right)^2\biggr).
\end{equation} 
         Thus, when more than one photon candidate was 
         obtained from a single electromagnetic cluster,
         candidates that had small energy
         compared to the cluster energy were rejected if their 
         reconstructed position
         was close  to the cluster centre. 
         The 40 mrad
         in the exponential is the average polar aperture 
         of a lead glass block 
         as seen from the interaction vertex. 
        The factor A was determined empirically
        from the Monte Carlo sample. It was set at 
        a value which ensured that the number of rejected 
        photons was less than 1/10 of the number of rejected 
        fake photons in all energy bins of $\rm E_{\gamma}. $
   \item A candidate was also rejected if it was found to
          satisfy:
\begin{equation}
        \rm  \theta_{trk} < B + C\cdot 
           exp\biggl(-\left({\frac{E_{\gamma}}{E_{trk}}}\right)^2\biggr).
\end{equation} 
        A photon candidate was likely to be fake if it had
        an energy, $\rm E_{\gamma}$, smaller than  
       the electromagnetic energy  which could
       be attributed to tracks.
       The factors B and C were determined  from the Monte Carlo
       sample. They were also set
       at values which ensured that the number of rejected 
       photons was less than 1/10 of the number of rejected 
       fake photons in all energy bins of $\rm E_{\gamma}. $
  \end{enumerate}
 The number of remaining fake photons was further reduced 
 using a weight function 
 $$\rm W(E_{\gamma},E_{clust}, \theta_{clust}, \theta_{trk}, E_{trk}) $$ 
 which was calculated for every photon candidate.
 The five variables were assumed uncorrelated and a likelihood
 ratio distribution was determined for each variable.
 The likelihood ratio in each bin was defined as the ratio
 of the number of 
 generated photons to the total number of photon candidates.
 The value of W was the product of the five likelihood ratios 
 of the bins 
 $\rm E_{\gamma},E_{clust}, \theta_{clust}, \theta_{trk} $ and
 $\rm E_{trk}$ in which the candidate was found.
 The discriminating power of W is shown in Figure
 \ref{fig:Purity}, where the efficiency 
 and purity for photons in Monte Carlo events 
 are shown as a function of a value 
 $\rm W_{cut}.$  
 For photon candidates with  $\rm W>W_{cut}, $ 
 the purity
 was defined as the ratio of the number of generated photons to the
 total number of reconstructed photons 
 and the efficiency 
 was defined as the fraction of generated photons which were correctly
 reconstructed.\par
 All possible pairs of photon candidates were then 
 considered. Each pair was assigned a probability P for both 
 candidates being correctly reconstructed photons, 
 the probability being simply the
 product of the weights W associated to the two candidates:
\begin{equation}
      \rm P  = W_1 \times W_2,\ with\ no \ cut\  on\ W_1\ or\ W_2.
\end{equation} 
 The combinatorial background consisted of a mixture of three
 components: wrong pairing of two 
 correctly reconstructed  photons, pairing of two fake photons 
 and pairing of one correctly reconstructed photon with 
 one fake photon. Choosing only photon pairs with high
 values of P reduces the combinatorial
 background to its  ``wrong pairing of correctly
 reconstructed photons'' component
 only. It was found that requiring  
 $\rm P>0.1 $ removed $\rm 60\% $ of the total combinatorial background,
 with a relative loss in efficiency of 
 $\rm 8\% $ and  $\rm 1\% $ 
 for $ \pi^0 $ and $ \eta $ signals, respectively.   
 The two-photon invariant mass distribution was studied in intervals 
 of the energy fraction $\rm x_E= \frac{E_{2\gamma}}{E_{beam}}. $ 
 Because most of the true reconstructed photons came from $ \pi^0 $
 decays,   
 an additional cut was required, in addition to the cut on 
 the probability P, to enhance 
 the $ \eta\rightarrow 2\gamma $ signal. 
 This cut excluded,  
 for invariant masses $\rm M_{2\gamma} >$ $\rm 300\ MeV, $
 any photon that could pair with any other photon
 to make an invariant mass $\rm M_{2\gamma} < $ $\rm 300\ MeV $ 
 with  a probability $\rm P > 0.1 $.\par
\begin{figure}
        \centerline{\includegraphics[scale=0.8]{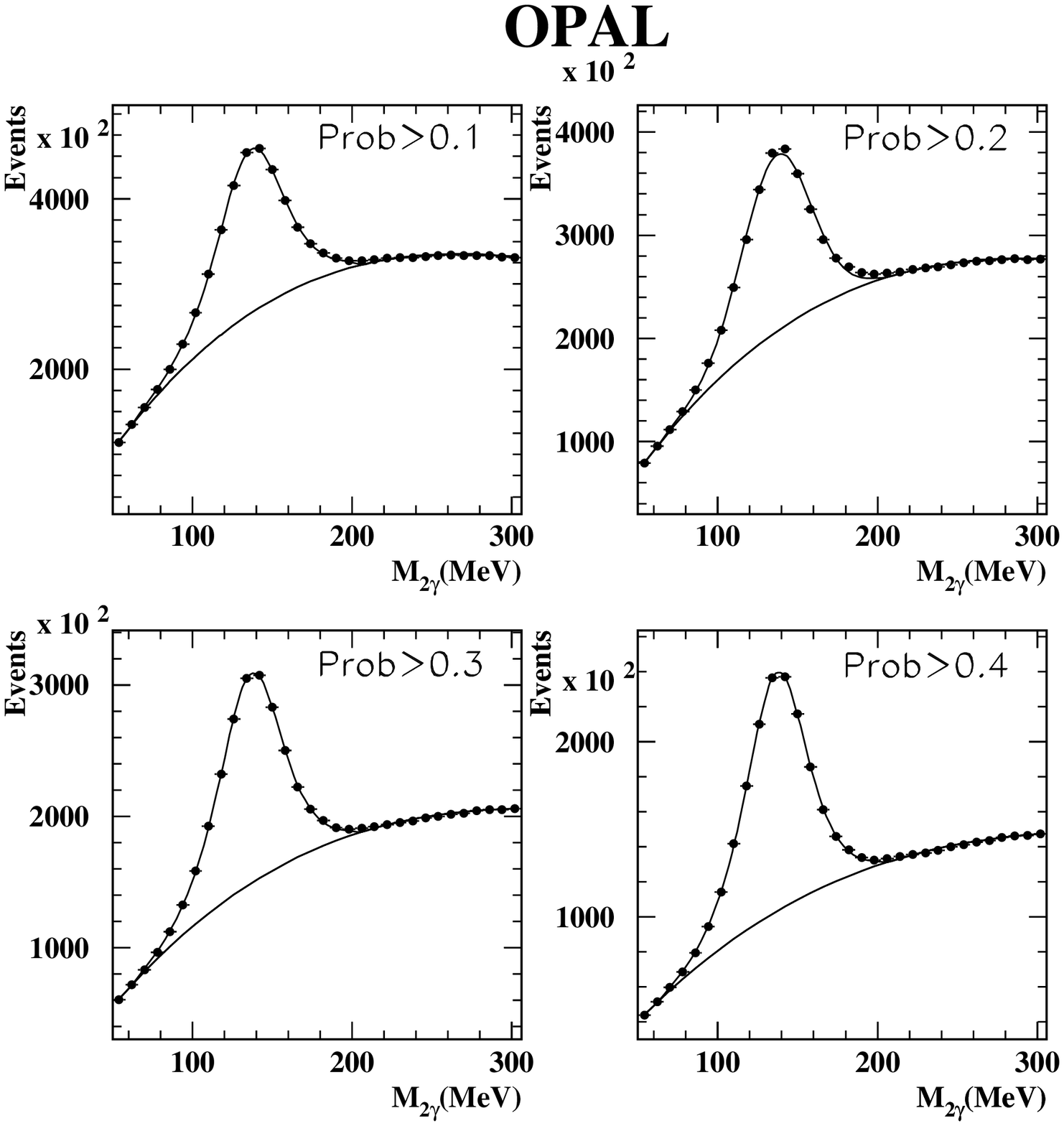}}
        \caption{Two-photon invariant   
         mass distribution in the vicinity
         of the $ \pi^0 $  for different cuts on 
         the probability (prob) $\rm P= W_1\times W_2$. 
         The smooth curve is from the
         fit to a double Gaussian and a second order
         polynomial. Note the suppressed zero on the vertical axis.}  
        \label{fig:pi0mass}
\end{figure}
\begin{figure}
        \centerline{\includegraphics[scale=0.8]{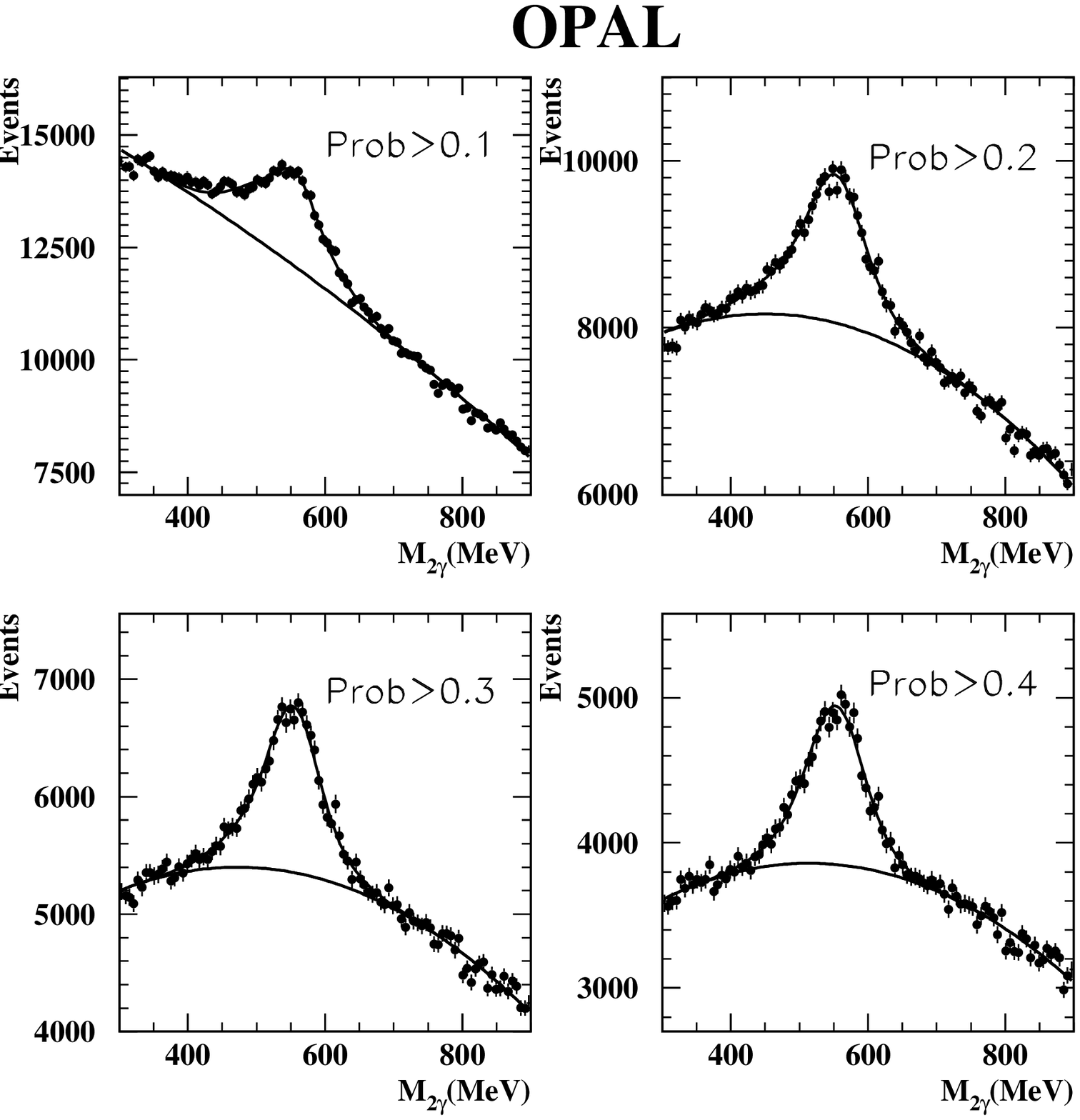}}
        \caption{Two-photon invariant   
         mass distribution in the vicinity
         of the $ \eta $  for different cuts on the probability 
         (prob) $\rm P= W_1\times W_2$. 
         The smooth curve is from the fit 
         to a double Gaussian and a second order
         polynomial. Note the suppressed zero on the vertical axis.} 
        \label{fig:etamass}
\end{figure}

 The combinatorial background could be described 
 by a second order polynomial  
 for all cuts on probability P, and for all $\rm x_E $ bins.
 The signal was well described by a double Gaussian.
 The mass distributions obtained from the data 
 over the full $\rm x_E $ range are 
 shown for four different cuts on the probability P
 in Figure \ref{fig:pi0mass}
   for  $ \pi^0, $ and in Figure \ref{fig:etamass} for $ \eta $.
   The absolute $ \pi^0 $ and $ \eta $ reconstruction 
   efficiencies, for $\rm P>0.1 $ over the entire
   $\rm x_E $ range, were $\rm 15.5 \% $ and 
   $\rm 7\% $ respectively. The efficiency and the
   signal to background ratio were dependent on the 
   $\rm x_E$ interval considered.
%  (see ~\cite{TN527}).  
   \par 
\subsection{ $\rm  K^0_S $ reconstruction }
\begin{figure}
        \centerline{\includegraphics[scale=0.8]{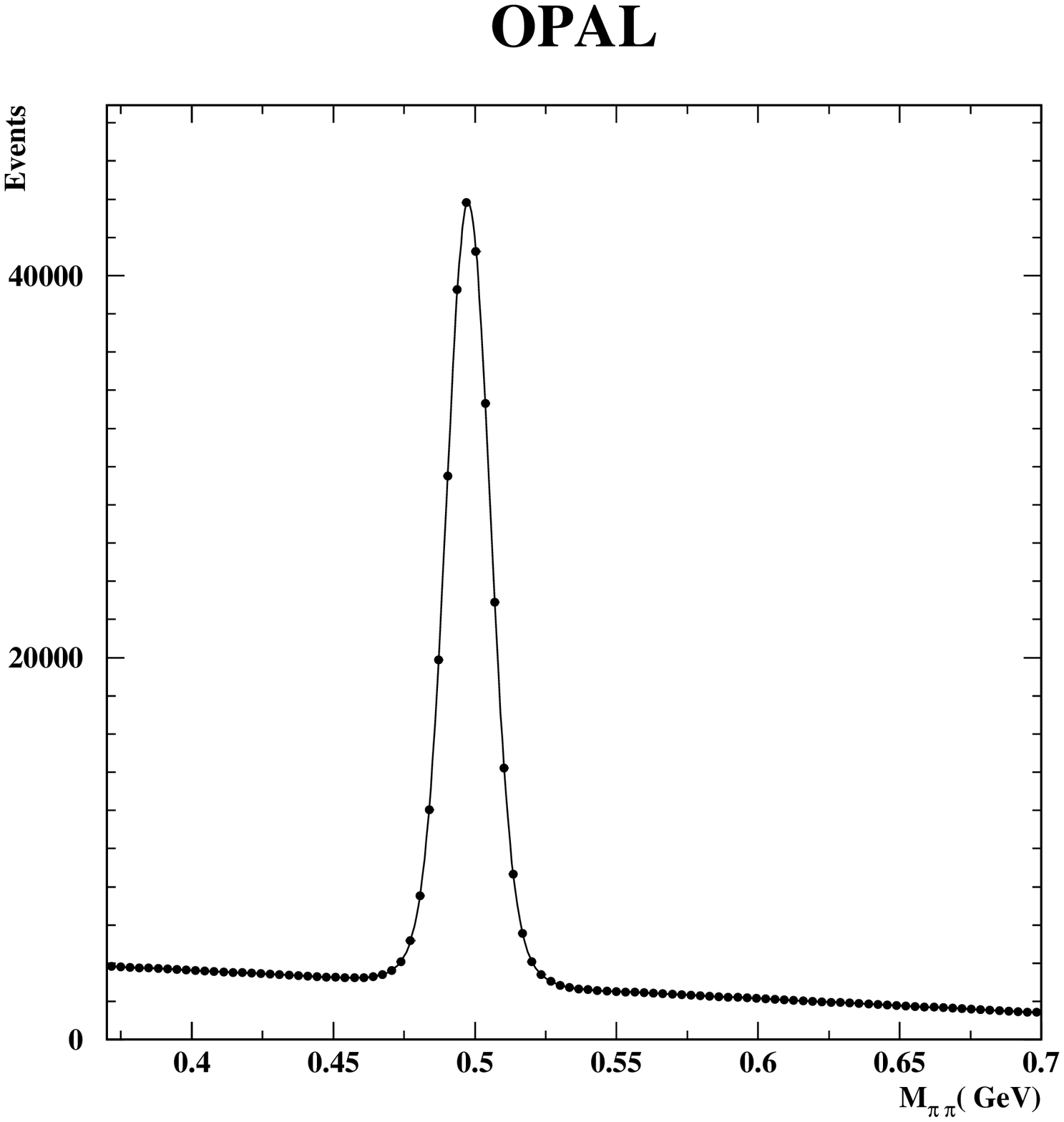}}
        \caption{$ \pi^+ \pi^- $   
         invariant mass distribution in the vicinity
         of the $\rm K^0 $. 
         The smooth curve is from the fit 
         to a double Gaussian and a second order
         polynomial.} 
        \label{fig:k0mass}
\end{figure}

 The $\rm K^0_S\rightarrow $ $  \pi^+\pi^- $ reconstruction was similar to the
 method described in~\cite{kaons}. Here it was applied to the
 full available $\rm Z^0 $ hadronic sample of four million events.
 A track was considered to be a pion candidate if 
 it had a transverse momentum larger than 150 MeV, had more than 20 hits
 in the jet chamber and had either more than 3 hits in the Z-chamber or a 
 reconstructed end point in the jet chamber~\cite{cj}.
 The invariant mass of pion pairs was evaluated for pairs of oppositely
 charged tracks 
 having an intersection point in the plane perpendicular to the beam axis
 and satisfying the following requirements:
\begin{enumerate}
\item the distance from the intersection point to the primary vertex had to 
      be greater than 1 cm and less than 150 cm;
\item if the secondary vertex was reconstructed in the jet chamber, it had
      to be less than 5 cm from the first hit of either track;
\item if the intersection point occurred before the jet chamber, 
      the radial distance from the track to the beam axis at 
      the point of closest approach had to exceed 3 mm;
\item track pairs that passed the above cuts were re-fitted with 
      the constraint that they originated from a common vertex;
\item track pairs that satisfied the photon conversion hypothesis
      or $ \Lambda\rightarrow p\pi $ hypothesis were rejected.
\end{enumerate}
The $ \pi^+ \pi^- $ invariant mass spectrum is shown in 
Figure \ref{fig:k0mass} for the whole measured $\rm x_E $ region.  
The spectrum was studied in  $\rm x_E $ intervals. In each interval,
a double Gaussian shape for the signal and a second
order polynomial for the background were used to fit the 
$\rm M_{\pi^+\pi^-} $ spectrum. The $\rm K^0_S $ reconstruction efficiency
was found to be $\rm 26\% $ for $\rm x_E < 0.1 $, reducing
to $\rm 15\% $ at higher $\rm x_E$.
\subsection{Inclusive $ \mathbf{\pi^0,\ \eta} $ and 
            $\rm \mathbf K^0 $ rates }   
   The inclusive rate measurements were performed on 
   the whole hadronic sample without making any 3-jet 
   requirement. 
   For each of 
   the mesons $ \pi^0,\ \eta $ and $\rm K^0_S $, the number of 
   entries remaining after the combinatorial background subtraction
   was considered as the number of signal
   entries. 
   This was corrected for detector and
   reconstruction efficiencies and for the non-measured 
   decay channels using the Particle Data Group branching 
   ratios.
   The total rates were also corrected for the inaccessible $\rm x_E $
   regions. The corrections were performed using extrapolation with  
   the spectral shapes predicted by Jetset. 
   They amounted to 
   $\rm 11\%,\ 23\% $  and $\rm 0.6 \% $ of the total rates for 
   $ \pi^0,\ \eta $  and $\rm K^0 $ respectively.
   The $\rm K^0 $ rate was the $\rm K^0_S $ rate corrected for 
   the non-observed $\rm K^0_L$.
   The total inclusive measured rates per event were:
   \begin{center}
    $ \rm\langle n_{\pi^0}\rangle
      \rm   = \ 9.871 \pm 0.040(stat)  \pm 0.39(syst)$\\
     $\rm\langle n_{\eta}\rangle   = \ 1.076 \pm 0.090(stat) \pm 0.084(syst)$\\
     $\rm \langle n_{ K^0}\rangle  = 
     \ 2.016 \pm 0.003(stat)  \pm 0.052(syst).$\\
   \end{center}  
   The results  are in good agreement with the values previously
   measured at LEP~\cite{kaons,LEP}. 
   The reconstruction
   method described in Section 5.1 gave an improved photon purity
   and higher $\pi^0\rightarrow 2\gamma $ and $\eta\rightarrow 2\gamma $
   reconstruction efficiencies when compared to~\cite{TN527}. This led
   to a well controlled combinatorial background in the 
   two-photon invariant mass spectrum. Thus, better systematic uncertainties
   were obtained.
   The statistical error on the $\eta $ inclusive rate was larger than
   the error quoted in~\cite{TN527}
   because the $ \pi^+\pi^-\pi^0 $ channel was not included in the
   present analysis.
   Since the full available $\rm Z^0 $ hadronic sample was analysed, 
   the statistical error on the $\rm K^0 $ inclusive rate was improved
   when compared to the value quoted in~\cite{kaons}. However, the new
   systematic error was slightly larger, due to the
   inclusion of the systematic error on the Monte Carlo background 
   and signal shapes (Section 5.4 item 1 to 3).\par
\begin{figure}
        \centerline{\includegraphics[scale=0.8]{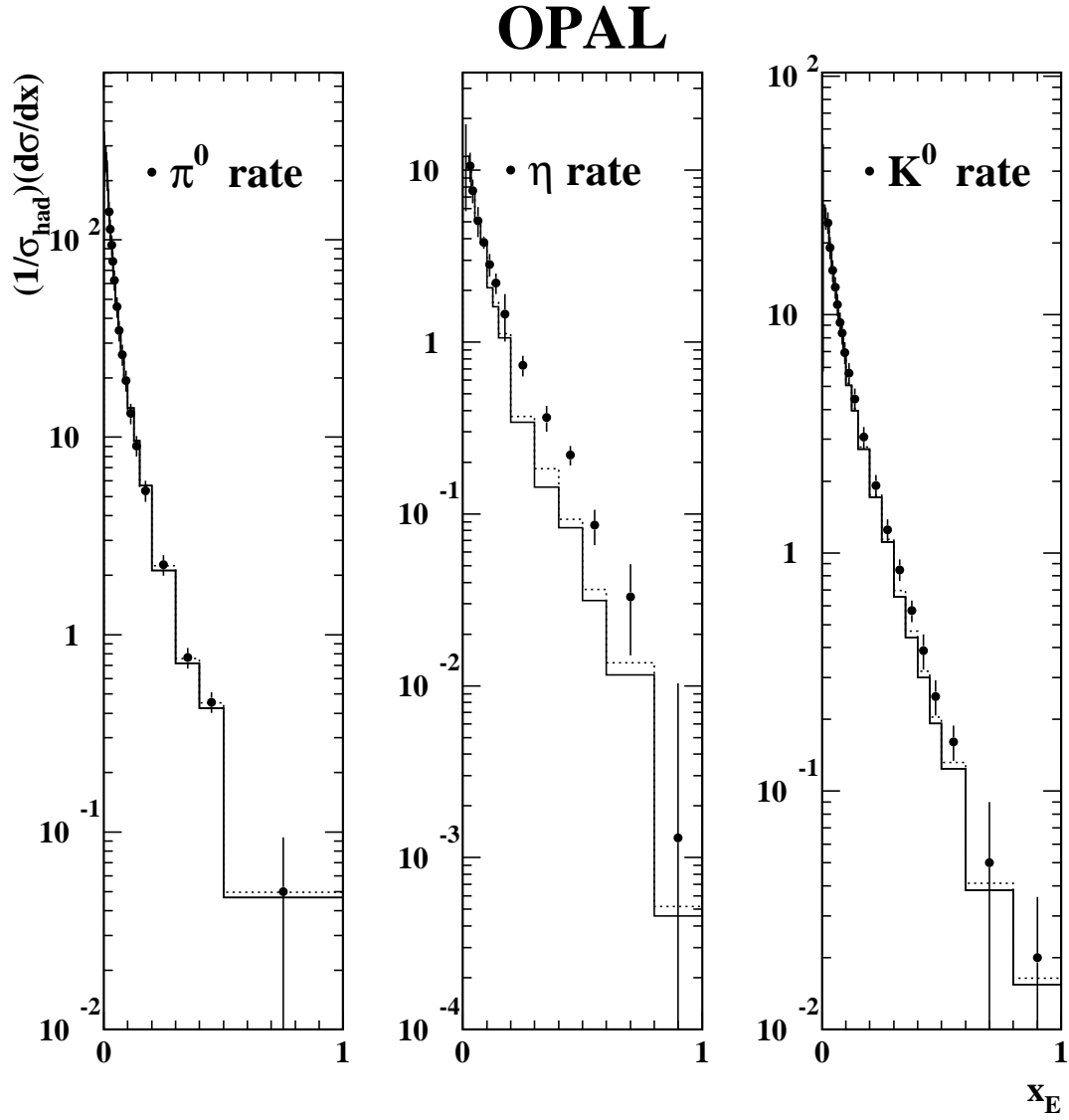}}
        \caption{ Differential rates of $ \pi^0, \eta $ and 
         $\rm K^0 $ in $\rm Z^0 $ hadronic decays. The errors for the data 
         points include the systematic errors. The histograms and the dotted
         lines are from  
         Jetset and Herwig Monte Carlo expectations respectively.} 
        \label{fig:full_frag}
\end{figure}

 
%It was inlcuded
%   for a more uniform treatment with $\pi^0$ and $\eta$ mesons.\par    
   The rate measurement was repeated  in $\rm x_E $ 
   intervals and the obtained differential rate distributions
   are shown in Figure \ref{fig:full_frag}.
   The $ \pi^0 $ measured differential rate 
   was described well by both Jetset and Herwig Monte Carlo
   expectations. 
   However, the measured $\eta $ and $\rm K^0 $ spectra were harder 
   than either Monte Carlo prediction. The discrepancy was worst 
   for $ \eta, $ where
   the measured rate was almost double the predicted rate for 
   $\rm x_E>0.2 $.  The $\rm K^0 $ measured rate was only about $\rm 15\% $ 
   larger than the predicted rate for $\rm x_E>0.2. $ 
   For $\rm x_E < 0.1 $ the  Monte Carlo
   predicted rates for $ \eta $ and $\rm K^0 $
   mesons agreed with the measured values within the 
   error bars.   
   These observations were in good agreement with previous
   OPAL results~\cite{TN527,kaons}. The derived values for
   $\rm \pi^0 $ and $\rm \eta $ 
   are not meant to supersede the former OPAL results.
   The new measured $\rm K^0 $ differential rates are given in Table 1.  
 \par 
\subsection{ Systematic errors }
   For the $\pi^0,\ \eta $ and $\rm K^0 $ 
   the largest contribution to the systematic error came from 
   the parametrisation of the 
   combinatorial background and the signal shape. This
   contributed up to $\rm 50 \% $ of the systematic error for
   some $\rm x_E $ intervals.   
   This error was  estimated using the following procedure for
   each $\rm x_E $ interval:
\begin{enumerate}
  \item The shape of the background was 
        measured from data by fitting a second order
        polynomial to the non-signal regions of the two-photon  
        invariant mass
        spectrum. The procedure was repeated using different side
        bands.
  \item The shape of the background was fixed to the shape 
        predicted by the Monte Carlo and the background
        was fitted to data allowing only an overall 
        normalisation factor to vary.
  \item The shape of the signal was fixed to the shape predicted by the 
        Monte Carlo and was
        fitted to data allowing only an overall 
        normalisation factor to vary.
 \end{enumerate}
        The systematic error was taken to be the quadratic sum 
        of all deviations from the value measured using only data.
        For the $\rm K^0 $ rate, the systematic errors in each 
        $\rm x_E $ interval were estimated using 
        the  procedure described in~\cite{kaons} to which 
        was added the contribution from the three
        items above.\par
        In the case of the $ \eta, $ the combinatorial 
        background was found to have 
        a small structure at $\rm M_{2\gamma}\simeq 700\ MeV $ in 
        the Monte Carlo which
        was not seen in the data. This structure was caused by  
        $  \omega(780)\rightarrow\pi^0\gamma\rightarrow\ 3\gamma$ for 
        which the rate in the OPAL-tuned Monte Carlo 
        was twice the measured rate.
        To estimate the systematic error the above procedure was repeated by:
\begin{enumerate}
       \item  fitting the background shape outside 
              the signal and the small structure regions;
       \item  removing from the Monte Carlo $\rm 50\% $
              of the generated $\omega(780)\rightarrow\ 3\gamma, $  
              in which case the structure disappeared.        
 \end{enumerate}
\par
     For both the $\rm \pi^0 $ and the $\rm \eta $ all the following systematic
     variations were considered:
\begin{enumerate}
\item      The systematic error relative to the cut on the probability 
      P associated with each photon pair was estimated by
      repeating the measurements for each 
      $\rm  x_E $ bin with different cuts on 
      $\rm P>0.2,\ P>0.3\ and\ P>0.4 $, the original value being obtained with
      $\rm P>0.1.$ This error, which took into account 
      signal purity and 
      reconstruction efficiency, since they depend on P, 
      was added quadratically to the previous error. In the worst
      case it contributed $\rm 14\% $ 
      to the total systematic error. 
\item      The difference between the correction   
      factors for detector effects obtained with Jetset and  
      those obtained with Herwig was considered as a systematic error. It was
      found to contribute up to  $\rm 12\% $ of the total systematic
      error.
\item      The error due to the energy calibration of the 
      electromagnetic calorimeter 
      was estimated from Monte Carlo by
      shifting up and down the energy of measured electromagnetic
      clusters by $\rm 2\%. $ This had
      a negligible effect on the number of reconstructed photon 
      candidates. However, the position of the peaks of the
      $ \pi^0 $ and
      $ \eta $ signals were shifted by about 10 MeV, and
      the difference in the signal extracted in each $\rm x_E $ bin
      was considered as a systematic error. In the worst case, this contributed
      $\rm 8\% $ to the total systematic error.
\item    Some of the mesons, mainly $ \pi^0, $ were produced in interactions
      with detector material. This effect might not have been very well
      modelled in the Monte Carlo. Therefore, half of the Monte Carlo
      prediction for these mesons was included in the uncertainty. 
      This accounted for up to $\rm 25 \% $ of the total systematic error, the
      worst being in the low $\rm x_E $ intervals. 
\item   An alternative extrapolation scheme as in~\cite{TN527} was
      used  to correct for the inaccessible $\rm x_E $
      regions. This yielded  
      slightly different corrections and these differences were
      included in the systematic error.
 \end{enumerate}
      \par 
\section{$ \mathbf \pi^0,\ \eta $ and $\rm \mathbf K^0$ production in 
             quark and gluon jets }
   The $ \pi^0 $, 
   $ \eta $ and $\rm K^0 $ yields in quark and gluon jets were 
   estimated as a 
   function of the scale $\rm Q_{jet}$.
   The average number of 
   mesons produced in the second jet and third jet 
   samples was measured as a function of $\rm Q_{jet} $, and then 
   the unfolding to $\rm 100\% $ quark and gluon jet purities was
   performed
   in the same way as was done for charged particles.\par
   Each $ \pi^0,\ \eta $ and $\rm  K^0 $ candidate 
   was assigned to the jet which made the smallest
   opening angle with respect to the total momentum
   direction of the meson. The 
   $ \pi^0,\ \eta $ and $\rm K^0 $ signals were then extracted as in the
   case of the inclusive rate measurement (see Section 5). 
   To make the jets fully comparable, and because photons were reconstructed
   only in the barrel region of the electromagnetic calorimeter, 
   only events that had both the second and third jet in 
   the barrel 
   were kept.
   The $\rm Q_{jet} $ interval was divided into three bins:
    $\rm Q_{jet} $ = 7 to 13,  13 to 19  and 19 to 25  GeV.
   The average number of mesons per jet, in each $\rm Q_{jet} $ interval, 
   was corrected for detector acceptance, reconstruction efficiency
   and for inaccessible $\rm x_E $ regions. 
   Efficiencies were calculated separately for each jet as a function 
   of $\rm Q_{jet} $.
   Using the average purity of each bin of $\rm Q_{jet} $ evaluated from 
   data using the matrix element formula, the average number of
   mesons per jet    
   was unfolded to $\rm 100\% $ quark and gluon jet purities.\\
\subsection{ $ \mathbf{\pi^0} $ production }
\begin{figure}
       \centerline{\includegraphics[scale=0.8]{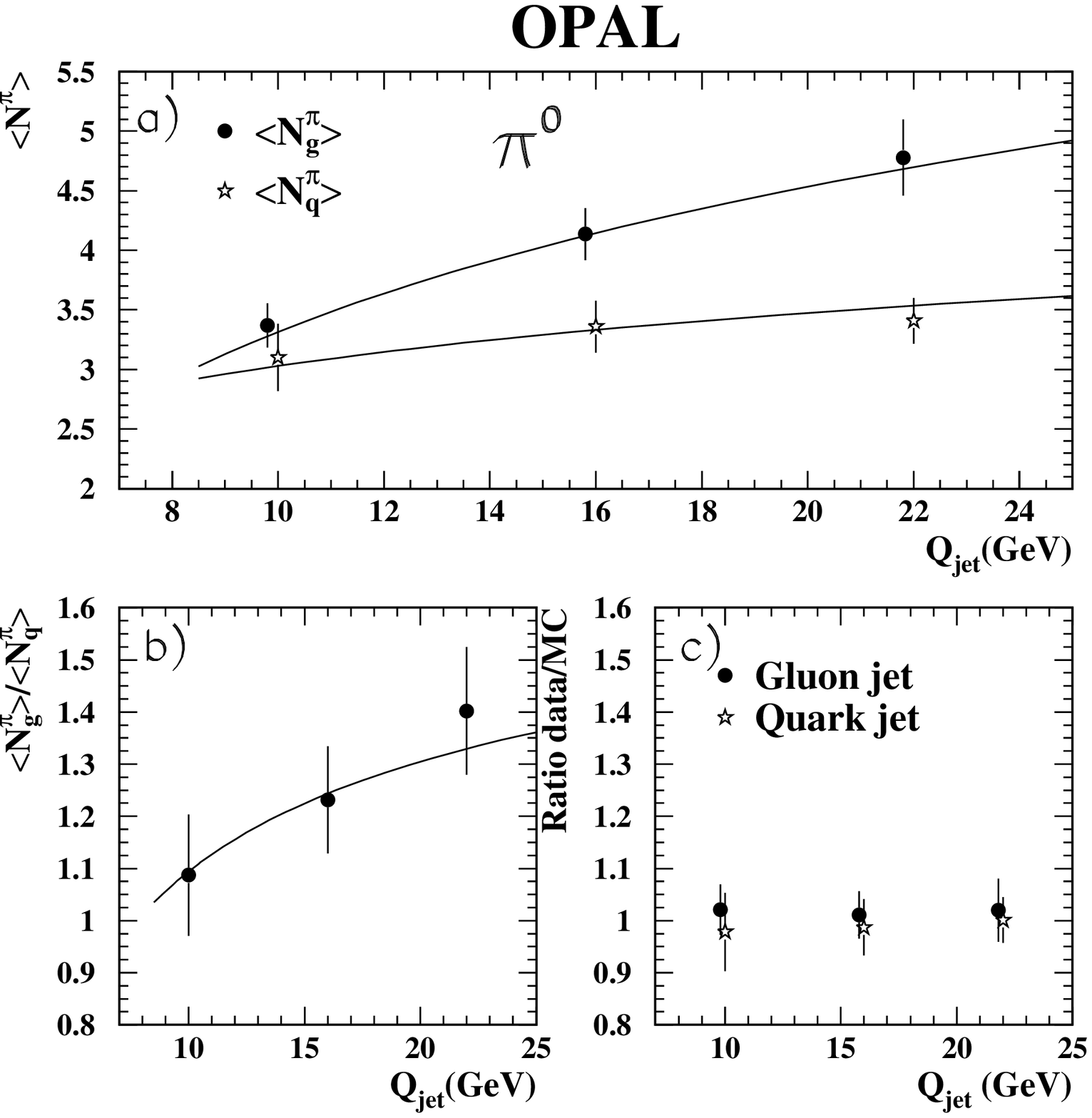}}
       \caption{ a) Average number of $ \pi^0 $ produced in 
        pure gluon and quark jets as a function of the scale $\rm Q_{jet} $. 
        The curves are the functions obtained for the 
        average charged particle multiplicity scaled by 
        a normalisation factor of 0.47.
        b) Ratio of production of $ \pi^0 $ in gluon and quark jets,
           as a function of $\rm Q_{jet} $.
           The curve 
           is the ratio of the functions obtained for
           the average charged particle multiplicity in gluon and quark jets. 
        c) Ratio of production rate of $ \pi^0 $ in the data to that 
           in the Jetset Monte Carlo, for gluon and quark jets. The gluon data
           points are shifted to the left for clarity.
          Error bars include systematic and 
          statistical errors added in quadrature.}  
      \label{fig:pi0_result}
\end{figure}

   The average number of $ \pi^0$ produced per jet as a function of 
   $\rm Q_{jet} $
   is shown in Figure \ref{fig:pi0_result}a for gluon and quark jets. 
   The ratio of multiplicities in gluon and quark jets
   is shown in 
   Figure \ref{fig:pi0_result}b.
   The Jetset and Herwig models were found to reproduce 
   the data within the error bars. This is demonstrated in Figure 
   \ref{fig:pi0_result}c where the ratios of data to Monte Carlo
   are compatible with unity for both quark and gluon jets in 
   each interval of $\rm Q_{jet}. $
   The analytical function obtained for the average charged particle 
   multiplicity as a function of $\rm Q_{jet} $, scaled with only one 
   free overall
   normalisation factor (measured to be 0.47), were found to fit well to
   the rate of $ \pi^0 $ as a function of $\rm Q_{jet} $  in gluon and quark
   jets, and to the ratio between $ \pi^0 $ 
   production rates in gluon and quark jets.
   The function is shown
   in Figures \ref{fig:pi0_result}a and \ref{fig:pi0_result}b.\par
\subsection{ $ \mathbf{\eta} $ production }
\begin{figure}
       \centerline{\includegraphics[scale=0.8]{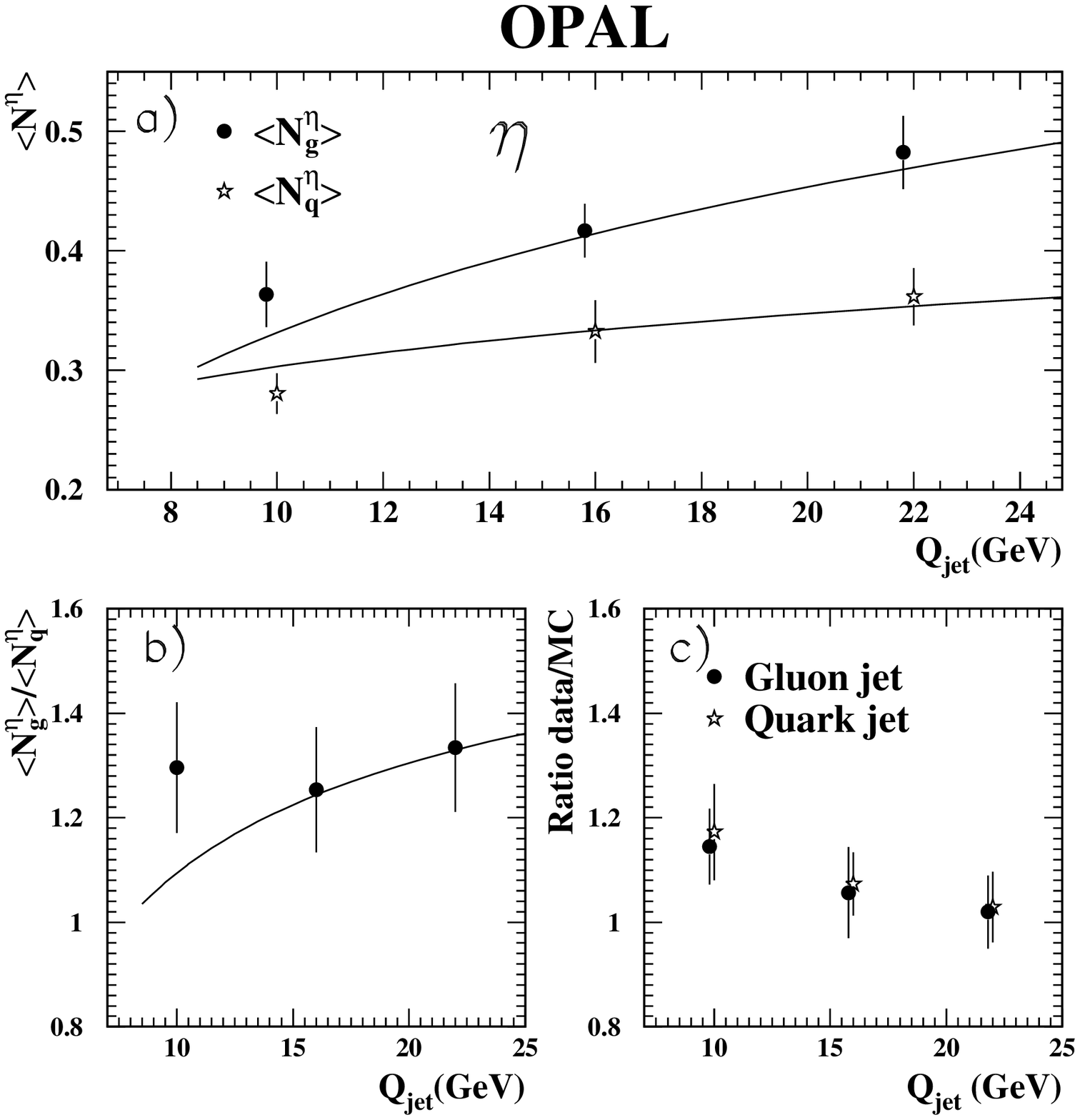}}
       \caption{ a) Average number of $ \eta $ produced in    
        pure gluon and quark jets as a function of the scale $\rm Q_{jet} $. 
        The curves are the functions obtained for the 
        average charged particle multiplicity scaled by 
        a normalisation factor of 0.047.
        b) Ratio of production of $ \eta $ in gluon and quark jets,
           as a function of $\rm Q_{jet} $.
           The curve 
           is the ratio of the functions obtained for
           the average charged particle multiplicity in gluon and quark jets. 
        c) Ratio of production rate of $ \eta $ in the data to that 
           in the Jetset Monte Carlo, for gluon and quark jets. The gluon data
           points are shifted to the left for clarity.
          Error bars include systematic and 
         statistical errors added in quadrature.}  
       \label{fig:eta_result}
\end{figure}

   The same analytical function, but with an overall normalisation factor
   of 0.047, was found to describe
   the $ \eta $ rate in gluon
   and quark jets, as shown in Figure \ref{fig:eta_result}a.
   The ratio of $ \eta$ multiplicities in gluon and quark jets as
   a function of $\rm Q_{jet} $
  is shown in Figure \ref{fig:eta_result}b, and was compatible
   with the ratio of charged  particle multiplicities. The ratio was also 
   compatible with being independent of $\rm Q_{jet} $, with  
    $$\rm \frac{ \langle N_g^{\eta} \rangle}
         { \langle N_q^{\eta} \rangle }
        = 1.29\pm 0.14. \quad  $$\par
   The measured $ \eta $ rate was found to be   
   slightly higher in the data than in the Monte Carlo,  
   mainly at low $\rm Q_{jet} $. This small disagreement was the same
   for both quark and gluon jets, as shown in 
   Figure \ref{fig:eta_result}c, where the ratios of data to Monte Carlo
   for gluon and quark jets are shown.
   This indicates that modelling of 
   production of $ \eta $ is equally inadequate for both 
   Jetset and Herwig Monte Carlos.
   No additional enhancement of 
   $\rm \eta $ production in gluon jets was 
   observed.\par
%\subsubsection{Dependence on the beam energy fraction $\rm x_E$ }
    Figure \ref{fig:full_frag}
    indicates  that the measured $\rm \eta$ spectrum
    was harder than the Monte Carlo prediction.
    To investigate this, the analysis was repeated for  
    $\rm  x_E \geq 0.1 $ for both $ \pi^0 $ and $ \eta $ mesons.
    Due to statistical limitations, mainly in the $ \eta$ meson sample,
    a yet harder cut on $\rm x_E $ was not appropriate.
    The resulting $ \pi^0 $ production rates 
    as a function of the scale $\rm Q_{jet} $
    for the gluon and the quark jets were in good agreement with both 
    Jetset and Herwig expectations.
    In each bin of $\rm Q_{jet} $, the ratio of data to Monte Carlo for the 
    $ \eta $ rate was equal for quark and gluon jets. 
    The ratio of the $ \eta $ production rates in data  
    to those predicted by the Monte Carlo were
    found to be the same for both the quark and gluon jets.
    The ratio of the production rate of 
    $ \eta $ in gluon jets to that in quark jets
    was also found to be consistent with being independent
    of $\rm Q_{jet} $.
    The measured ratio was:         
    $$\rm \frac{ \langle N_g^{\eta}(x_E>0.1)\rangle}
         { \langle N_q^{\eta}(x_E>0.1)\rangle}
        = 1.38\pm 0.19. \quad $$ 
     \par
%        = 1.38\pm 0.19 \quad .$$
    This result is not in contradiction with previously 
    published results~\cite{L3ETA} concerning an excess of 
    high momentum  $ \eta $ mesons over Monte Carlo prediction. The
    excess is present equally in 
    quark and gluon jets.\par
\subsection{ $\rm \mathbf{K^0} $ production }
\begin{figure}
       \centerline{\includegraphics[scale=0.8]{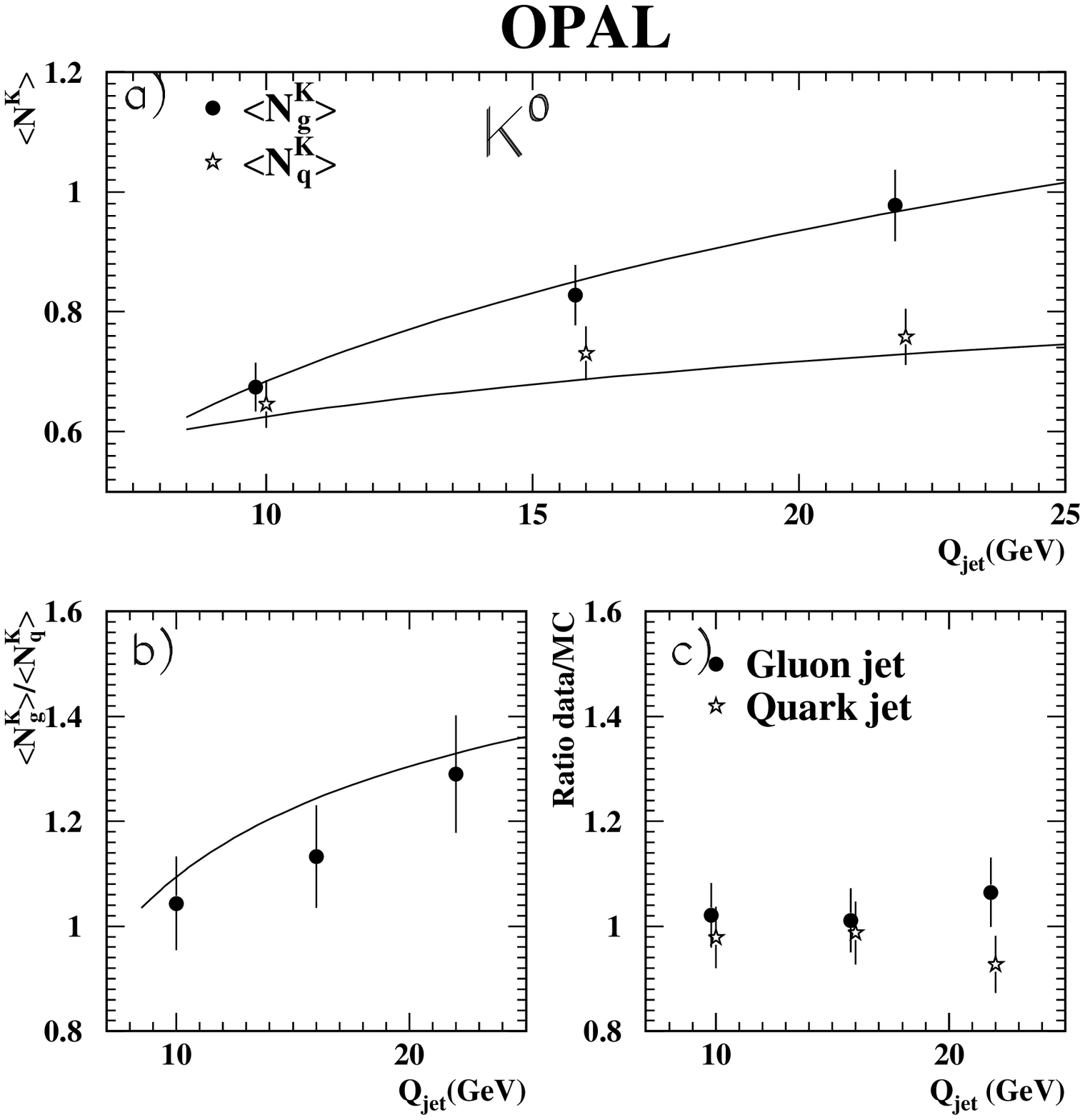}}
       \caption{ a) Average number of $\rm K^0 $ produced in 
        pure gluon and quark jets as a function of the scale $\rm Q_{jet} $. 
        The curves are the functions obtained for the 
        average charged particle multiplicity, scaled by 
        a normalisation factor of 0.097.
        b) Ratio of production of $\rm K^0 $ in gluon and quark jets,
           as a function of $\rm Q_{jet} $.
           The curve 
           is the ratio of the functions obtained for
           the average charged particle multiplicity in gluon and quark jets. 
        c) Ratio of production rate of $\rm K^0 $ in the data to that 
           in the Jetset Monte Carlo, for gluon and quark jets. The gluon data
           points are shifted to the left for clarity.
          Error bars include systematic and 
          statistical errors added in quadrature.}  
       \label{fig:k0_result}
\end{figure}

     An overall scale factor of 0.093 in 
    the analytical function of Section 4 
    gave the best fit to the $\rm K^0 $ production rate in quark and 
    gluon jets. However, as shown in 
    Figure \ref{fig:k0_result}a, the data for the
    quark jet were systematically but not significantly,
    higher than the corresponding 
    analytical shape. This is also shown in Figure \ref{fig:k0_result}b
    where the ratio of $\rm K^0 $ production in gluon and quark jets
    was systematically smaller than the corresponding ratio
    for charged particle production,
    although compatible within errors. This effect could be
    explained by a higher probability of producing a strange meson
    in the fragmentation of an initial strange, charmed or 
    b quark.\par
    The ratios data to Monte carlo of the $\rm K^0 $ production rate in
    gluon and quark jets are               
    shown in Figure \ref{fig:k0_result}c. The ratios
    are compatible with unity within errors and agree with 
    the previous OPAL result~\cite{LEPALL}.\par    
\subsection{ Systematic errors }
     The measurements of the production rates of
     $ \pi^0,\ \eta $ and $\rm K^0 $ in jets were not
     statistically limited. The errors shown in Figures 
     \ref{fig:pi0_result}, \ref{fig:eta_result} 
     and \ref{fig:k0_result} already include both 
     the systematic and statistical errors added in quadrature  and  
     propagated through the unfolding formula. The  
     systematic error ranged from $\rm 60\% $ to $\rm 90\% $ of the total 
     quoted error, depending on the data point. It was estimated 
     for each jet and each $\rm Q_{jet} $ interval using the procedure
     described in Section 6 for $ \pi^0 $, $ \eta $ and 
     $\rm K^0 $ inclusive rate measurements.
     In each $\rm Q_{jet} $ interval,
     the  selection criteria (see Section 5) for 
     $\rm \pi^0,\ \eta \ and \ K^0 $ mesons 
     were found to act equally on quark and gluon jets. 
     The results were very stable against changes of the
     selection cuts that caused relative 
     reconstruction efficiency variation of up to $\rm 5\%.$  
     For the $\rm \eta $ case,  
     the additional cut (Section 5.1) 
     that excluded for invariant masses $\rm M_{2\gamma} >$ $\rm 300\ MeV, $
     any photon that could pair with any other photon
     to make an invariant mass $\rm M_{2\gamma} < 300\ MeV $ 
     with  a probability $\rm P > 0.1 $, was changed: 
     first $\rm M_{2\gamma} < 300\ MeV $ was moved to 
     $\rm M_{2\gamma} < 200\ MeV $ and second  $\rm P > 0.1 $ was replaced 
     by $\rm P > 0.2. $ The effect on the $\rm \eta $ measured
     rates in quark and gluon jets was negligible($\rm <1\% $).  
     In addition, for all studied mesons: 
     \par
\begin{enumerate}
 \item The difference between the measurements with purities taken 
       from the matrix
       element formula and from Monte Carlo(see Section 3) 
       was negligible.
       The average jet topologies and purities were varied by changing 
       the cut on the variable $\rm Y $ (see Section 2.3)  
       such that the purities obtained with the matrix element formula
       still agreed with those obtained from 
       Monte Carlo information. The analysis was repeated
       with $\rm Y\le 0.25 $ and the differences obtained in each bin of
       $\rm Q_{jet} $ were considered as systematic errors and were found 
       to contribute at most $\rm 5\% $ of the 
       total quadratic sum.  
  \item  Since the mesons were reconstructed 
       independently of the jet-finding, and
       were assigned to jets by angular matching once the jets were 
       reconstructed, very little dependence on the 
       jet finder was expected. Indeed, this 
       was the case for Luclus and
       Durham jet finders where the difference was 
       measured to be less than $\rm 2\%$. However,
       the results obtained with the cone 
       jet finder showed deviations of up to
       $\rm 10\% $ with respect to the two other jet finders. This was 
       considered as a systematic error, and  
       contributed up
% here now the relative error increased to 47% instead of 40 % 
       to $\rm 47 \% $ of the total quadratic sum of the systematic errors.
\item  The results were found to be stable in each bin of $\rm Q_{jet} $
       with respect 
       to changes of the charged particle selection requirements. 
       The results were also found to be stable when the  
       number of charged particles required per 
       jet was increased from               
       3 to 5.
       Lowering the cut on the inter-jet angle from 30 to $\rm 20^\circ$ was 
       correlated to the cut on the variable Y. The analysis 
       was repeated with  the minimum inter-jet angle set to $\rm 20^\circ $ 
       and 
       the minimum value of Y set to 0.22. The maximum deviation obtained 
       was less than $\rm 2\% $ and was added to the total systematic error.
\item 
      Due to statistical limitations, the analysis 
      could not be repeated on data using b-tagging
      to obtain gluon enriched jet samples.
      A systematic error, that could account for
      correlations between particle content in 
      pure quark and pure gluon jets as well as for the stability
      of the unfolding method, was assigned using the following procedure: 
      The analysis was repeated        
      on four different Jetset Monte Carlo samples, each sample 
      being as large as the full available data sample.
      The jet samples were selected to have different 
      gluon and quark average purities. The purities  were set based 
      on Monte Carlo information (Section 3.2) to be:
      $\rm (40\%\ quark, 60\%\ gluon), $
      $\rm (25\%\ quark, 75\%\ gluon), $
      $\rm (60\%\ quark, 40\%\ gluon), $
      $\rm (75\%\ quark, 25\%\ gluon)  $
      for the four pairs of jet samples that were processed with 
      the unfolding method.
      The largest deviation with respect to
      the average unfolded meson rate 
      was considered as a systematic error.
      This contributed up to $\rm 7\% $ of the systematic error,
      indicating that the unfolding procedure was indeed stable
      against large purity variations. 
% Attention 
%( here now the error is decreased to 7% instead of 15%)
%      It was 
%      not the dominant systematic error, as was the case        
%      for charged particles.
\end{enumerate}       
\section{ Summary and conclusion }
       Average multiplicities of $ \pi^0 $, $ \eta $,
       $\rm K^0 $ and 
       charged particles have been  measured for  gluon and quark jets 
       as a function of a transverse momentum-like scale $\rm Q_{jet} $. 
       The average multiplicities 
       were unfolded to $\rm 100\% $ purity  by comparing two jet samples 
       having different gluon (or quark) content for the same value 
       of $\rm Q_{jet}$.
       The $ \pi^0,\ \eta$ and $\rm K^0 $ 
       inclusive production rates 
       in $\rm Z^0 $ hadronic decays were measured  and found to agree
       with the previous values measured at LEP. 
       The $ \pi^0 $ production rate was found to 
       be well described by both Jetset and
       Herwig Monte Carlos for all values of $\rm x_E$.
       The $ \eta $ and $\rm K^0 $ spectra
       were found to be harder than  
       the Monte Carlo predictions, particularly in the case 
       of $\rm\eta $.\par 
       The charged particle multiplicity 
       as a function
       of the  topological scale $\rm Q_{jet} $
       for pure quark and pure gluon jets was
       described by a simple phenomenological 
       formula. The same formula, with all parameters fixed except for
       an overall normalisation factor, was found to provide 
       a good fit to the $ \pi^0, $ $\rm \eta $ and $\rm K^0$
       production rates in 
       gluon and quark jets. 
       The analysis  showed 
       that for $ \pi^0,\ \eta $ and $\rm K^0 $ mesons, there was
       no evidence for an enhancement in gluon jets 
       with respect to quark jets beyond the enhancement observed
       for  inclusive charged particles.
       In particular we observed no evidence for an enhancement
       of $\rm \eta $ meson production in gluon jets, contrary
       to the predictions of some models for gluon jet
       hadronisation.\par
       We measured the ratio of the slope 
       of the average charged particle multiplicity 
       in gluon jets to that in quark jets. We obtained 
       $\rm C=2.227\pm 0.07(stat.)\pm 0.19(syst.)$ for 
       this ratio, that is about    
       one standard deviation of the total uncertainty above 
       the analytic next-to-next-to-next-to leading order (3NLO) 
       prediction~\cite{Capella}.\par 
%=======================================================================
\section{Acknowledgements}
\label{s:acknowledgements}

We particularly wish to thank the SL Division for the efficient operation
of the LEP accelerator at all energies
 and for their continuing close cooperation with
our experimental group.  We thank our colleagues from CEA, DAPNIA/SPP,
CE-Saclay for their efforts over the years on the time-of-flight and trigger
systems which we continue to use.  In addition to the support staff at our own
institutions we are pleased to acknowledge the  \\
Department of Energy, USA, \\
National Science Foundation, USA, \\
Particle Physics and Astronomy Research Council, UK, \\
Natural Sciences and Engineering Research Council, Canada, \\
Israel Science Foundation, administered by the Israel
Academy of Science and Humanities, \\
Minerva Gesellschaft, \\
Benoziyo Center for High Energy Physics,\\
Japanese Ministry of Education, Science and Culture (the
Monbusho) and a grant under the Monbusho International
Science Research Program,\\
Japanese Society for the Promotion of Science (JSPS),\\
German Israeli Bi-national Science Foundation (GIF), \\
Bundesministerium f\"ur Bildung und Forschung, Germany, \\
National Research Council of Canada, \\
Research Corporation, USA,\\
Hungarian Foundation for Scientific Research, OTKA T-029328, 
T023793 and OTKA F-023259.\\

%=======================================================================

\vfill\eject
\begin{table}
   \begin{center}
    \begin{tabular}{|c|c|}
       \hline
       $ \rm x_E  \ range $ & $\rm \frac{1}{\sigma_{had}} \times
       \frac{d\sigma}{dx_{E}} $ \\ 
       \hline
       \hline
  0.0114 - 0.020 & $\rm  25.731\pm    0.232\pm   1.430 $\\ 
  0.020 - 0.030  & $\rm  24.617\pm    0.120\pm   1.300 $\\ 
  0.030 - 0.040  & $\rm  19.349\pm    0.116\pm   1.040 $\\ 
  0.040 -  0.050  & $\rm  15.500\pm    0.061\pm   0.767 $\\  
  0.050 -  0.060  & $\rm  13.170\pm    0.072\pm   0.690 $\\ 
  0.060 -  0.070  & $\rm  11.144\pm    0.073\pm   0.600 $\\  
  0.070 -  0.080  & $\rm   9.360\pm    0.066\pm   0.500 $\\
  0.080 -  0.900  & $\rm   8.470\pm    0.061\pm   0.468 $\\
  0.090 -  0.100  & $\rm   7.010\pm    0.059\pm   0.401 $\\
  0.100 - 0.125  & $\rm   5.734\pm    0.029\pm   0.312 $\\ 
  0.125 - 0.150  & $\rm   4.488\pm    0.028\pm   0.247 $\\
  0.150 - 0.200  & $\rm  3.100\pm    0.019\pm    0.169 $\\
  0.200 - 0.250  & $\rm  1.945\pm    0.015\pm    0.104 $\\
  0.250 - 0.300  & $\rm  1.266\pm    0.010\pm    0.071 $\\ 
  0.300 - 0.350  & $\rm  0.860\pm    0.010\pm    0.050 $\\ 
  0.350 - 0.400  & $\rm  0.579\pm    0.009\pm    0.035 $\\
  0.400 - 0.450  & $\rm  0.394\pm    0.008\pm    0.026 $\\ 
  0.450 - 0.500  & $\rm  0.253\pm    0.005\pm    0.018 $\\ 
  0.500 - 0.600  & $\rm  0.163\pm    0.003\pm    0.018 $\\ 
  0.600 - 0.800  & $\rm  0.051\pm    0.001\pm    0.010$\\ 
       \hline       
       \end{tabular}\label{tab:k0table}
     \end{center}
\caption{$\rm K^0 $ differential 
 rate.}
%%%%% in  hadronic $\rm Z^0 $ decays.}
\end{table}
\end{document}